\documentclass[sigconf]{acmart}
\AtBeginDocument{%
  \providecommand\BibTeX{{%
    \normalfont B\kern-0.5em{\scshape i\kern-0.25em b}\kern-0.8em\TeX}}}

\copyrightyear{2021} 
\acmYear{2021} 
\setcopyright{acmcopyright}
\acmConference[CHI '21]{CHI Conference on Human Factors in Computing Systems}{May 8--13, 2021}{Yokohama, Japan}
\acmBooktitle{CHI Conference on Human Factors in Computing Systems (CHI '21), May 8--13, 2021, Yokohama, Japan}
\acmPrice{15.00}
\acmDOI{10.1145/3411764.3445381}
\acmISBN{978-1-4503-8096-6/21/05}

\usepackage{lipsum}
\usepackage{xargs} 
\usepackage{xcolor}
\usepackage{tikz}
\usepackage{soul}
\usetikzlibrary{shapes}
\usepackage{enumitem}
\usepackage{paralist}
\usepackage{hyperref}
\newcommandx{\updated}[2][1=]{}

\DeclareRobustCommand{\colorHLine}[1]{\tikz \draw [line width=0.5mm, {#1}] (0,0) -- (.8em,0);}
\DeclareRobustCommand{\colorVLine}[1]{\tikz \draw [line width=0.5mm, {#1}] (0,0) -- (0,.7em);}
\definecolor{colorAbstractTask}{HTML}{417505}

\DeclareRobustCommand{\colorRect}[1]{\tikz \draw [fill={#1},draw=none] (0,0) -- (0.4,0) -- (0.4,0.15) -- (0,0.15) -- cycle;}

\definecolor{frameworkColor}{HTML}{D2D2D2}
\definecolor{colorLegendBlue}{HTML}{B9DEE8}
\definecolor{colorLegendYellow}{HTML}{F2E091}
\definecolor{colorLegendNYTLevel2}{HTML}{F9C467}
\definecolor{colorLegendNYTLevel3}{HTML}{FFA93D} 
\definecolor{colorLegendNYTLevel4}{HTML}{FF8B23}  
\definecolor{colorLegendNYTLevel5}{HTML}{FC6A0D} 
\definecolor{colorLegendNYTLevel6}{HTML}{F04F0A} 
\definecolor{colorLegendNYTLevel6}{HTML}{F04F0A}  
\definecolor{colorLegendNYTLevel7}{HTML}{E13006} 
\definecolor{colorLegendNYTLevel8}{HTML}{CE0706}
\definecolor{colorLegendLightOrange}{HTML}{FFC072}
\definecolor{colorLegendOrange}{HTML}{FE6E0D}
\definecolor{colorLegendRed}{HTML}{CE0706}
\definecolor{colorLegendLightGrey}{HTML}{F2F2F2}

\DeclareRobustCommand{\colorHLine}[1]{\tikz \draw [line width=0.4mm, {#1}] (0,0) -- (.6em,0);}
\DeclareRobustCommand{\colorVLine}[1]{\tikz \draw [line width=0.4mm, {#1}] (0,0) -- (0,.5em);}
\definecolor{colorFramework}{HTML}{5D5D5D}

\definecolor{JaniceEdit}{HTML}{FF2727}

\newcommand{\jz}[1]{#1}

\begin{document}

\title{Mapping the Landscape of COVID-19 Crisis Visualizations}

\author{Yixuan Zhang}  
\affiliation{\institution{Georgia Institute of Technology}}
\email{yixuan@gatech.edu}

\author{Yifan Sun}  
\affiliation{\institution{William \& Mary}}
\email{ysun25@wm.edu}

\author{Lace Padilla}  
\affiliation{\institution{University of California Merced}}
\email{lace.padilla@ucmerced.edu}

\author{Sumit Barua}  
\affiliation{\institution{Northeastern University}}
\email{barua.s@northeastern.edu}

\author{Enrico Bertini}  
\affiliation{\institution{New York University}}
\email{enrico.bertini@nyu.edu}

\author{Andrea G. Parker}  
\affiliation{\institution{Georgia Institute of Technology}}
\email{andrea@cc.gatech.edu}

\renewcommand{\shortauthors}{Zhang et al.}

\begin{abstract}
In response to COVID-19, a vast number of visualizations have been created to communicate information to the public. Information exposure in a public health crisis can impact people’s attitudes towards and responses to the crisis and risks, and ultimately the trajectory of a pandemic. As such, there is a need for work that documents, organizes, and investigates what COVID-19 visualizations have been presented to the public. We address this gap through an analysis of 668 COVID-19 visualizations. We present our findings through a conceptual framework derived from our analysis, that examines who, (uses) what data, (to communicate) what messages, in what form, under what circumstances in the context of COVID-19 crisis visualizations. We provide a set of factors to be considered within each component of the framework. We conclude with directions for future crisis visualization research. 
\end{abstract}



\maketitle

\section{Introduction}
The COVID-19 pandemic is touching many aspects of human life. Information has been rapidly generated to inform the public about the pandemic, help people understand complicated mathematical forecasting models, and persuade them to make behavioral changes to mitigate the spread of the disease. In addition to all the textual information being disseminated about COVID-19, a significant number of visualizations have been created to help people make sense of the complexities of the pandemic and understand the evolving public health crisis. \textit{Crisis visualizations} (i.e., visual representations of crisis information such as disease prevalence, epidemiological simulations, and economic and social changes) are being circulated daily. COVID-19 visualizations have been produced by diverse content creators (e.g., scientists, government and healthcare officials, social media users, news media outlets) and disseminated to a large number of audiences. This pandemic may be the first time in history that such a large proportion of the public has been engaged with and responding to visualizations.    

Research has shown that visualizations are useful to help the general public, who often have difficulties understanding complex crisis dynamics, making sense of crisis situations, assessing personal risk, and  decision-making~\cite{stone2015effects}. Prior work has suggested that visualizing information helps reduce mental load and allows people to grasp content more quickly, compared with presenting that information in text~\cite{dur2014interactive}. Research has examined crisis visualization design from various perspectives, such as providing guidance on how to design risk maps effectively~\cite{dent2008thematic, fostermapping, roth2012visualizing} and visualizing uncertainty to communicate risks~\cite{dieckmann2015home, dieckmann2017seeing, frewer2002public, padilla2017effects, van2019communicating}. Moreover, there has been an ongoing discussion among researchers, practitioners, and the public regarding whether or not to create COVID-19 visualizations~\cite{Correll_medium} and how to design these visualizations responsibly~\cite{Makulec_medium}.   

Given this proliferation of visualization production and consumption, there is an urgent and critical need to document and organize these efforts in a timely manner in order to provide guidance for emerging visualizations. Moreover, it is important to understand what visualizations have been designed and disseminated to the public because information consumption can impact people's behaviors, attitudes, and thus ultimately the path of the pandemic. Unfortunately, to date, there has not been a review of crisis visualizations that aim to communicate with the general public from a visualization perspective.  
 
To address this research gap, we curated and analyzed 668 visualizations that communicate information about COVID-19. Through our analysis, we derived a conceptual framework that focuses on understanding \textbf{who}, uses \textbf{what data}, to communicate \textbf{what messages}, in \textbf{what form}, as well as emphasizing the \textbf{contexts (what circumstances)} in which crisis visualizations are created. 
In this paper, we use this framework to discuss several characteristics of COVID-19 visualizations. First, we describe the variety of stakeholders who created the visualizations in our corpus. Second, we examine the characteristics of data within these visualization---including the data sources, data source citations, data quality and uncertainty. Third, we provide an explication of the six categories of messages inherent within these visualizations: informing of severity, forecasting trends and influences, explaining the nature of the crisis, guiding risk mitigation, communicating risk, vulnerability, and equity, and gauging the multifaceted impacts of the crisis. Fourth, we unpack the trends in the visualization techniques and encodings that have been used to communicate each type of message. Lastly, we explore the dynamic temporal context of crisis visualization design, specifically, how COVID-19 visualizations have changed over time. 

Our work makes contributions to visualization, crisis informatics, and public health research and practice. 
First, our findings will help to accelerate the design process of ongoing and future crisis visualizations by outlining currently used visualization methods and visual encoding techniques. By outlining these trends as well as open challenges, our findings can help inspire new visualization approaches and research areas and help designers avoid common pitfalls, thus accelerating innovation. 
Second, our work contributes to crisis informatics research, a field of study that examines the role of information and communication technologies in times of crisis. Little work in this field has examined the nature of visualizations designed for the general public during times of crisis. Through our comprehensive analysis of crisis visualizations, we contribute new knowledge regarding the nature of such visualizations. Specifically, we provide a systematic synthesis of crisis visualizations that characterizes trends and patterns in who, communicated what messages, to whom, in the dynamic temporal context of the COVID-19 pandemic. In addition, by formalizing our findings around the COVID-19 crisis visualization landscape in a preliminary framework \jz{---by extending and blending existing models in communication research and visualization research---we provide a thinking tool that aims to descriptively capture dimensions of the crisis visualization design space and important areas that warrant further inquiry.}
Third, through our framework and accompanying findings that document the kinds of messaging inherent in COVID-19 visualizations, this work will help catalyze research at the intersection of visualization and public health that examines the impact of consuming COVID-19 visualizations on behaviors and attitudes. Specifically, we provide a discussion of ways in which future work can further examine questions around the creation and impact of crisis visualizations raised in each area of our framework. Our framework will provide researchers with a conceptual tool that can guide empirical investigation, by highlighting important areas of inquiry in the crisis visualization space (e.g., investigating how exposure to different categories of crisis visualizations deferentially impacts risk perceptions, prevention behaviors, and mental health). 

\section{Related Work}
\label{sec:background}

\subsection{Crisis Informatics} 
\label{sec:crisis_informatics}

Traditional risk and crisis communication has emphasized a one-way, top-down approach to information creation and dissemination (e.g., transmitting information from entities such as governmental authorities to the general public~\cite{hagar2010introduction, zhang2020ci}). Today, much risk and crisis communication has shifted to two-way information exchange between the institutions and governmental authorities and the general public, a shift which necessitates considerations of the needs and values of each stakeholder group. For example, social media platforms enable direct communication between the authorities and the public in various forms (e.g., following, mentioning, replying)~\cite{kim2018emergency, zhang2019social}, which enables two-way communication. 
The involvement of the general public in crisis communication has transformed information and communications technology design to be more participatory~\cite{zhang2019social}, a paradigm shift that has helped fuel the development of the crisis informatics research area. 
 
\textit{Crisis informatics} is an interdisciplinary area that examines the interconnectedness of people, organizations, information, and technology during crises~\cite{hagar2010introduction}. Crisis informatics research has studied how crisis and risk communication has occurred online~\cite{chauhan2017providing}, and the use of information and communications technology during crises~\cite{zhang2020ci}. 
Despite the increasing interest in crisis informatics research, less work has examined the creation and dissemination of visualizations in times of crisis. Notable examples of work in this area include research examining ``public reporting'' and ``crisis mapping''---this research has examined participatory maps and crowdsourced crisis databases, and built computational models to provide risk warning and mitigation approaches~\cite{herranz2014multi, norheim2010crowdsourcing}. Another body of work has focused on the design and development of visualization tools to help people understand and explore trends arising during emergency situations and unfolding crises ~\cite{onorati2019social, romero2016towards}. However, overall, there has been little work that explores the intersection of visualization and crisis informatics. We address this gap, identifying challenges and opportunities in this field through a comprehensive analysis of crisis visualizations, such as how public engagement may shape the design of crisis visualization.

\subsection{Crisis Visualizations}
\label{sec:crisis_vis}

With increases in the amount of information about crises and associated risks has come an increase in the production and dissemination of visual representations of this information. \textit{Risk visualization} can be defined as ``the systematic effort of using images to augment the quality of risk communication along the entire risk management cycle''~\cite{roth2012visualizing}. Risk visualizations aim to depict the risks that people face and to help institutions and authorities communicate with the general public about risks. More recently, data collection, analysis, modeling, and visualizations of outbreak data has become increasingly complex, leading to an emerging field of \textit{outbreak analytics}, where visualization plays an important role in supporting sensemaking of complex outbreak data~\cite{polonsky2019outbreak}. The target audiences for visualizations of outbreak analytics mainly include health professionals and analysts. Visualizations in these fields may be defined distinctly depending on the contexts and approaches, the concepts are clearly interconnected---they are visual representations that aim to augment communication about undesirable situations that have inherent uncertainty, and that posit threats to human beings~\cite{zhang2020ci}. As such, we define \textit{crisis visualizations} as visual representations of information about undesirable situations that posit threats to human beings. These visualizations communicate information such as disease prevalence, epidemiological simulations, and economic and social changes. We will be using the term \textit{crisis visualizations} throughout the paper.

Prior work has explored data visualizations of historical pandemic events~\cite{lu2004web, preim2020survey, welhausen2015visualizing} \jz{and other crises, such as hurricane~\cite{Bica2019Com, padilla2017effects}}. This body of work has examined how visualizations influence risk perception~\cite{welhausen2015visualizing} and explored the design of visual analytic tools for modeling and simulation~\cite{polonsky2019outbreak}. A more recent survey on visual analytics for public health~\cite{preim2020survey} has presented a variety of requirements and visual analytics techniques to help guide epidemiologists and environmental health specialists conduct visual analyses of public health issues. There is thus a significant body of work that has focused on crisis visualizations that assume specialized audiences and require higher levels of numeracy and visualization literacy to interpret them. In contrast, our work examines crisis visualizations that aim to communicate with the general public. 

To date, research focused on COVID-19 visualization has been limited. Several notable exceptions include recently published work by D’Ignazio and Klein~\cite{dignazio2020datafeminism}, as well as Bowe, Simmons, and Mattern~\cite{bowe2020learning}. Both works have provided critical consideration of broader issues, such as equity, power, and social justice associated with COVID-19 data. Bowe et al.~\cite{bowe2020learning} used the lens of ``embodiment'' to interpret COVID-19 visualizations and discussed the importance of creating more meaningful visualizations through the use of embodied and local data. 
 
However, to our knowledge, existing work has not reported on syntheses of large crisis visualization datasets, and thus may not sufficiently depict a broad picture of COVID-19 visualizations. Considering that these crisis visualizations have received huge public engagement and the significant role of visualization in communicating crisis information, comprehensive and systematic analyses of COVID-19 visualizations are needed. Our work aims to identify emerging trends and patterns in currently used visualization methods and visual encoding techniques for COVID-19 crisis visualizations, as well as to identify challenges and common pitfalls.

\section{Methodology}
\label{sec:methodology}
This paper reports on our analysis of a COVID-19 visualization collection that we assembled. To compile this collection, we used opportunistic sampling, also called emergent sampling, a non-probability sampling method for data collection. Opportunistic sampling ``takes advantage of unforeseen opportunities after fieldwork has begun,'' and is particularly useful for synthesizing exploratory research areas~\cite{patton1990qualitative}. The time-sensitive and unpredictable evolution of the COVID-19 pandemic can make a priori sampling decisions challenging, thus motivating our use of opportunistic sampling, an approach that enabled us to flexibly collect visualizations as they were produced and disseminated. Several other works in crisis informatics have successfully used opportunistic sampling to study a variety of crisis events before (e.g., ~\cite{kassam2006mental, tang2020need}). However, we also need to point out the limitations of opportunistic sampling. Opportunistic sampling can suffer from selection bias and generalizability issues for findings, necessitating care when interpreting the results.

\subsection{Data Collection}
Our collection includes visual representations that communicate information about COVID-19, including both data visualizations (e.g., an interactive map, graph, chart, or diagram) and infographics (e.g., a static narrative or graphics). At the beginning of the data collection, early March 2020, we collected an initial set of visualizations. Later in March, to expand the collection, we began compiling a corpus of crisis visualizations focused on COVID-19 using image database searches (e.g., Google Images) and word-of-mouth contributions. We also searched on visualization blogs that contained special topics on the pandemic, such as \texttt{coronavirus} on FlowingData and \texttt{Science \& Health} on FiveThirtyEight. We then cleaned the collection by removing duplicated entries, so that each entry in the database had a unique URL. Whenever possible, we included the original visualization that appeared online in our database. Also, we removed entries with broken URLs (websites that were not maintained) on July 31, 2020. Note that some visualizations in our collection have been updated regularly; therefore our analysis reflects the state of these visualizations at the time we visited the website.

\subsection{Corpus}

\begin{figure*}[tb]
\centering
\includegraphics[width=\textwidth]{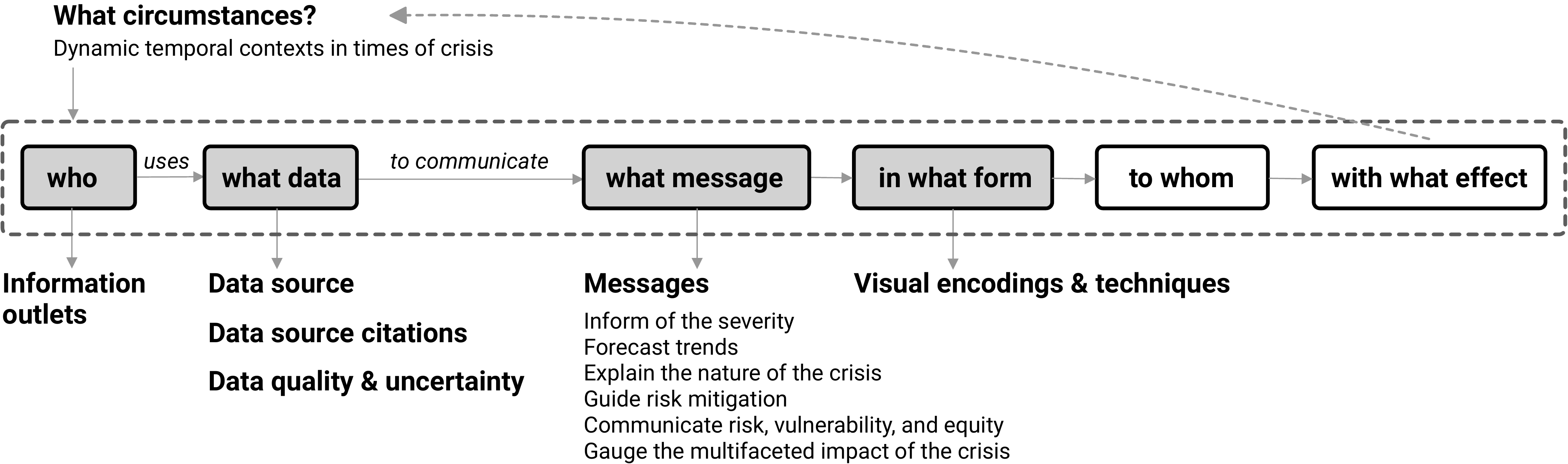}
\caption{A conceptual framework for understanding crisis visualizations, built upon existing models in communication research~\cite{braddock1958extension, lasswell1948structure} and visualization research~\cite{munzner2009nested}, focuses on \textit{who}, (uses) \textit{what data}, (to communicate) \textit{what message}, \textit{in what form}, highlighted in gray~\colorRect{frameworkColor}. 
Our framework also emphasizes the role of circumstances and contexts in understanding crisis visualizations, indicated by the surrounded dashed lines \colorVLine{colorFramework} \colorHLine{colorFramework} \colorHLine{colorFramework} \colorHLine{colorFramework}  \colorVLine{colorFramework}. 
The back dashed arrow $\dashleftarrow$ indicates a non-linear flow of visualization design and communication.
} 
\label{fig:framework}
\end{figure*}

To focus the scope of this work, we analyzed a subset of visualizations from our total collection of 1184 visualizations. The inclusion criteria for this work include visualizations in English that were published online and aimed to communicate with the general public. We used information outlets as a proxy for the target audience (i.e., general public vs. specialized audiences), for example, including visualizations shared on news media and governmental websites that provide information for the general public. We did not include academic publications, which typically assume specialized audiences. With these criteria in mind, this paper reports on our analysis of 668 visualizations (published between January 22 and July 31, 2020).  

\subsection{Analysis}

To analyze our visualization corpus, we engaged in a two-step process that involved both inductive and deductive coding. This multi-phase process enabled a rigorous and systematic analysis of our data set. We first conducted an inductive analysis~\cite{thomas2006general} of the collected visualizations, to develop an initial set of codes that were grounded in the phenomena reflected within our corpus. At the beginning of our analysis in late April (when we collected our initial set of visualizations), two researchers first individually reviewed our entire collection to develop an initial understanding of the visualization themes arising in the data. The two researchers then inductively coded the first 100 visualizations in the collection, and then clustered related low-level codes to arrive at high-level themes that characterize the visualizations. The two researchers discussed the coding scheme regularly to review the evolving codes to achieve a mutual understanding and to refine the codebook. The codebook contains a table that has a list of the codes and their definitions that reflect the concept referred to by the code. The goal of the codebook was to maximize coherence among coders~\cite{creswell2017research}.
We also consulted with senior visualization researchers and had conversations with visualization practitioners and journalists who created COVID-19 visualizations to further refine the codebook, which allowed us to add new codes and reorganize the codes. This multiphase process enabled us to arrive at a detailed and multifaceted codebook with which to characterize the COVID-19 visualizations in our corpus.

Our final codebook included 61 codes within four categories: metadata (e.g., title, URL, publisher), type of visualizations  (e.g., choropleth map), messages that the visualizations aim to communicate (e.g., informing of the severity), data handling such as whether the visualization used normalized data. Guided by the codebook, we then conducted a deductive analysis of the visualizations~\cite{thomas2006general}, which enabled us to exhaustively assess to what extent the various concepts in our codebook were reflected across our entire visualization corpus. We re-coded the previously-coded 100 visualizations, and continued coding the rest of the visualizations in the collection at that time. In each phase of coding, the researchers discussed any points of disagreement in the application of codes until we reached a consensus. The codebook and the corpus used for this paper are provided as supplemental material and also available at \href{https://osf.io/gf8zr/}{https://osf.io/gf8zr/}.

\begin{figure*}[tb]
\centering
\includegraphics[width=\textwidth]{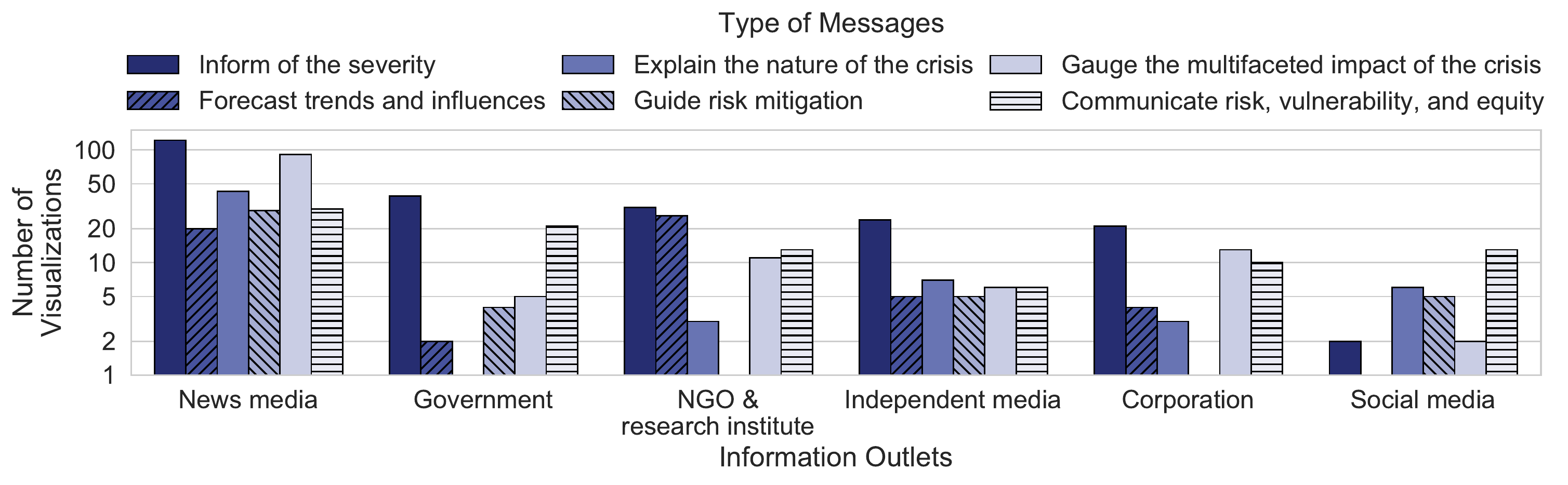}
\caption{\jz{The distribution of COVID-19 visualizations in our corpus, by information outlets and type of messages communicated (the y-axis is in a logarithmic scale).}} 
\label{fig:message_outlet}
\end{figure*}

\section{A Conceptual Framework for Crisis Visualizations}
\label{sec:unique_characteristics}

Through our analysis, we derived a conceptual framework for crisis visualizations (see ~\autoref{fig:framework}). Our framework focuses on \textbf{who}, uses \textbf{what data}, communicates \textbf{what messages} in \textbf{what form}, and emphasizes the \textbf{contexts (what circumstances)} in which visualizations are created, based on Lasswell’s model of communication~\cite{lasswell1948structure}, Braddock's extended formulation of Lasswell's model~\cite{braddock1958extension}\jz{, and Munzner's nested model~\cite{munzner2009nested}}.  Lasswell's model of communication (1940)~\cite{lasswell1948structure} describes an act of communication by defining ``Who, said what, in which channel, to whom, with what effect?''. Critiques of the model include that it neglects the impact of environments on communication. Braddock (1958)~\cite{braddock1958extension} thus advocated the addition of ``under what circumstances’’ to the model, to emphasize the attributes of time and contexts, and their influences on the communication process. \jz{The nested model by Munzner~\cite{munzner2009nested} highlights domain situations, task and data abstractions, visual encoding, and algorithms. In line with this nested model, our proposed framework also emphasizes domain situations and data and visual encoding. The difference lies in the prescriptive versus the descriptive nature of the work. Munzner’s nested model ``provides prescriptive guidance for determining appropriate evaluation approaches by identifying threats to validity unique to each level''~\cite{munzner2009nested}. Our framework is descriptive in the sense of providing concepts and guiding further inquiry through organizing and analyzing surveyed visualizations to capture their commonalities. Moreover, the nested model does not explicitly focus on ``who'' (i.e., content producers), ``communicates what messages'', ``to whom'', ``with what effect''.} Building upon \jz{these existing models in communication and visualization research,} our conceptual framework aims to support the analysis of crisis visualization communication patterns, \jz{to capture dimensions of the crisis visualization design space, and highlight areas that are worth further inquiry.}

In our framework, the first component is \textit{who} is creating COVID-19 visualizations (briefly described in \autoref{sec:who}). The second component is \textit{what data} is used for creating visualizations  (\autoref{sec:data}), which focuses on: a) understanding the diversity of data sources in crisis visualizations, b) data source citation practices, and c) practices of dealing with uncertainty caused by unavailable, missing, incomplete, and inconsistent crisis data. The third component is \textit{what message} is being communicated through existing visualizations (\autoref{sec:messages}). The fourth component is \textit{in what form} is data presented in visualizations (\autoref{sec:visual_forms}). The fifth component examines  \textit{under what circumstances} visualizations are created (\autoref{sec:contexts}), with a particular focus on understanding the dynamic temporal nature crises. 

As a preview, our findings suggest that crisis visualizations have been changed over time to fit the constantly changing crisis situations. Moreover, the impacts that visualizations make on people (e.g., shifting attitudes and public perceptions) then, in turn, impact how and what is being designed, and thus the process of design and communication is non-linear~(as reflected by the back dashed arrow $\dashleftarrow$). 
While our findings do not focus on the final two components of the framework---(\textit{to whom}) and \textit{with what effect}---our framework and the insights gained from our analysis do suggest important areas for future inquiry into these components. We discuss such opportunities in the Discussion section (\autoref{sec:discussion}).

\section{Who created the visualizations?}
\label{sec:who}

In our collection, there were 174 distinct information outlets that shared COVID-19 visualizations. Most visualizations were shared in news media outlets (57\%, 380 out of 668, e.g., the New York Times), followed by non-governmental organizations (NGOs) and research institutes (14\%, 95 out of 668, e.g., Johns Hopkins University), governmental agencies and public health officials (9\%, 60 out of 668, e.g., United States Centers for Disease Control \& Prevention), independent media outlets (8\%, 52 out of 668, e.g., FlowingData), corporations (7\%, 51 out of 668, e.g., McKinsey \& Company), and social media posts (4\%, 27 out of 668, e.g., Twitter). \autoref{fig:message_outlet} summarizes the different messages focused on by each information outlet in our corpus. We will describe this messaging in more detail in~\autoref{sec:messages}.

\jz{
In terms of the visualization country of origin, the majority of the surveyed visualizations in our corpus are from the United States of America (66\%, 439 out of 668), followed by the United Kingdom (20\%, 131 out of 668), Canada (3\%, 18 out of 668), India (2\%, 16 out of 668). The rest 9\% belong to Switzerland (n=8), Japan (n=6), China (n=6), Germany (n=6), New Zealand (n=5), Belgium (n=4), Iraq (n=4), Singapore (n=3), South Africa (n=3), Qatar (n=3), Italy (n=2), Australia (n=2), Finland (n=2), Sweden (n=1), Portugal (n=1), Brazil (n=1), Iceland (n=1), Bangladesh (n=1), Nigeria (n=1), France (n=1), Denmark (n=1), and for two visualizations, we were unable to determine the country of origin. }

\section{What data has been used to create the visualizations?}
\label{sec:data}

In this section, we provide an overview of data sources described in our database, the current status of data source citations, and data source quality and uncertainty. 

\subsection{Data Sources} 

We identified 486 distinct data sources that were used to communicate various aspects of the pandemic; among which 200 data sources were used to create visualizations to convey the message of the severity of the pandemic (see ~\autoref{sec:inform_severity}). Recent work has discussed how available COVID-19 data has varied in format, which in turn has led to inconsistent and incomplete data reporting with varied quality~\cite{lau2020internationally}. We also see the sheer variety of data sources represented in our corpus. To help investigate data quality issues, groups of ``digital volunteers'' have self-organized---online communities of volunteers across fields and cultures who collaboratively collect, verify, and map crisis information across digital channels. For example, the team of 1Point3Acres~\cite{1Point3Acres}---a group of volunteers with different backgrounds such as computer science, public policy, public health, and biology---have engaged in fact-checking and cross-referenced different sources in an effort to produce more accurate COVID-19 data.

We also identified eight visualizations that used crowdsourced data sources. These visualizations were all from research institutes and NGOs. The crowdsourced data has been used to help understand activity in communities, such as examining additional characteristics of the pandemic, early detection, and local forecasting of the pandemic. In our corpus, for example, COVIDcast~\cite{cmu_covid} includes multiple crowdsourced indicators, such as doctor visits and hospital admissions (based on data from health system partners), self-reported symptoms through surveys (advertised by Facebook),  search trends (data from Google), and COVID-19 antigen tests (data from Quidel). The resulting visualization was a choropleth map that integrated various crowdsourced indicators. Similarly, COVIDNearYou~\cite{covid_near_you} also applies crowdsourcing methodologies to bootstrap public health tracking (e.g., asking about website visitors' health status daily) as a way of identifying current and potential illness hotspots.

\subsection{Data Source Citations} 
 
Including data source citations in visualizations is one of the most important ways of supporting data provenance. Data provenance refers to data source citations, references, methodological choices,  relevant facts, and annotations of exceptions and correction~\cite{hullman2011visualization}.  93\% of visualizations in our collection cited the original data source, suggesting that the majority of the visualizations were rigorous in this respect. However, we still see that approximately 7\% did not include source citations, which undermines the credibility of those visualizations.
Among the uncited visualizations, 32\% were used to depict the severity of the pandemic, followed by communicating risks, vulnerability, and equity (24\%), gauging the impact of the pandemic (17\%), guiding risk mitigation (10\%), and explaining coronavirus (5\%). Some of these visualizations were screenshots shared on social media (e.g., Twitter, Reddit) without source citations. These visualizations went viral on the internet~\cite{viralgraph2020}, potentially spreading and bolstering fear and uncertainty, and driving harmful reactions such as taking medicine that has not been fully tested and not taking the pandemic seriously. Therefore, having data sources cited is important to improve the validity of the visualizations. 

\subsection{Data Source Quality and Uncertainty} 

Currently unavailable, missing, incomplete, and inconsistent information may cause uncertainty. Our analysis shows that COVID-19 visualization designers communicated uncertainty related to data quality in three ways: adding disclaimers, including explanations of the data collection (e.g., describing missing data or inconsistencies in reporting), and assessing the uncertainty caused by the data quality issues. For example, in our corpus, the COVID Tracking project~\cite{covid_tracking_project} offers data-quality grades, which assess how well each US state reports COVID-19 data. The grade is based on 16 separate factors within 5 categories, including reporting (how well states format and publish data), testing data completeness (whether a state is publishing complete basic testing data), patient outcomes (whether a state is reporting on COVID-19's effect on patients), demographics (whether a state reports the break-down information by demographic categories), and other (e.g., whether a state reports hospital capacity). Attempts such as these may help provide the public with guidelines for interpreting and formulating decisions based on data that has inherent uncertainty.

\section{What messages do the visualizations communicate?}
\label{sec:messages}


COVID-19 crisis visualizations communicate a wide range of information about the pandemic. Based on our analysis, we classify these messages as reflecting six purposes: (1) informing of the severity; (2) forecasting trends and influences; (3) explaining the nature of the crisis; (4) guiding risk mitigation; (5) communicating risk, vulnerability, and equity; and (6) gauging the multifaceted impacts of the crisis. Note that these message categories are not mutually exclusive. In our corpus, it is common for one visualization to convey more than one message. 

\textbf{(1) Informing of the severity:} 
The severity of a pandemic, as represented by the number of diagnoses, hospitalized patients, incubated patients, and deaths, is one essential aspect of public health communication~\cite{hastall2013severity} and crisis communication~\cite{cori2017key}. The importance of communicating the severity of a crisis is reflected by the large proportion of the visualizations in our database (231 out of 668, 35\% of the visualizations aim to communicate the severity of the pandemic). Indeed, this was the message category most represented in our corpus. By far, most visualizations were created to inform the public about the current state of the crisis, including the trajectory of the crisis over time and the geospatial spread of the crisis (218 out of 231, 94\%). A much smaller proportion (13 out of 231, 6\%) compared COVID-19 with historical events. 
Most of the visualizations that compared with historical events (8 out of 13, 62\%) in our collection were published in the early stage of the outbreak (January to March 2020). One reason for this early focus on historical events may be that there was a greater level of uncertainty during the initial phase of the crisis, and as such comparing the pandemic with historical events may have been seen as a way to help the public make sense of what was happening.  

\textbf{(2) Forecasting trends and influences:}  
Visualizations have been created to estimate and forecast trends and influences in terms of estimation of current transmissibility, effective reproduction rate, and projections of the pandemic (57 out of 668, 9\%). 

\textbf{(3) Explaining the nature of the crisis:}
A body of visualizations in our corpus used biomedical information to explain the nature of the COVID-19 crisis (95 out of 668, 14\%), including the structure of the virus (25 out of 95, 26\%), illness symptoms and timeline of diagnosis (26 out of 95, 27\%), how disease transmission happens (31 out of 95, 33\%), and approaches to contact tracing (13 out of 95, 14\%). 

\textbf{(4) Guiding risk mitigation:}
Visualizations have been also created to guide risk mitigation (46 out of 668, 7\%), including providing practical instructional guidance to keep social distancing, how to prepare and use personal protective equipment (PPE), how to maintain proper hygiene, \jz{as well as justification for particular risk mitigation strategies}. 
 
\begin{figure*}[tb]
\centering
\includegraphics[width=\textwidth]{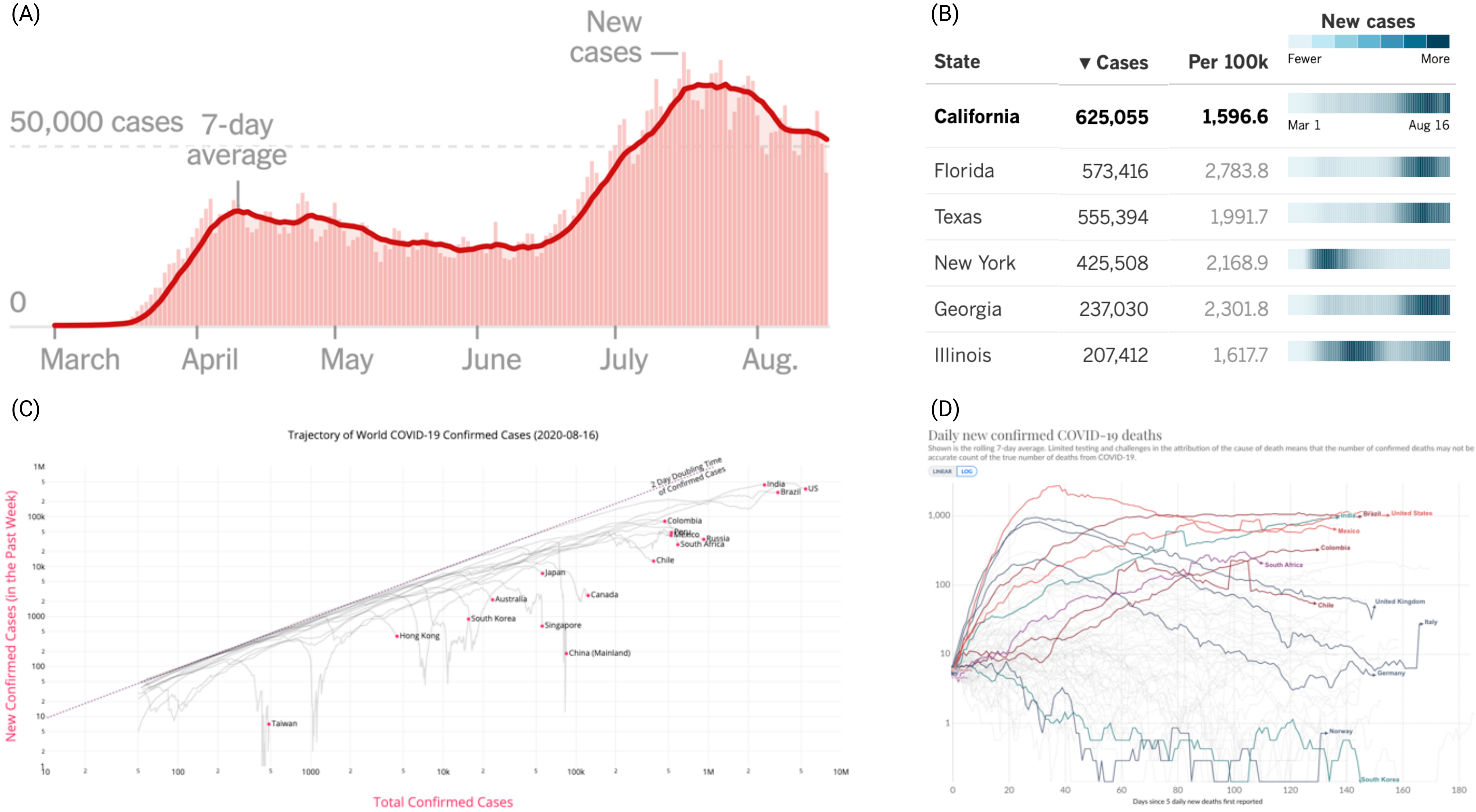}
\caption{Examples of temporal visualizations: 
(A) New reported cases by day in the United States, superimposed with a 7-day average line (by the New York Times~\cite{nyt_dashboard}); 
(B) A set of Pez Charts showing case number changes over time (by the Los Angeles Times~\cite{latimes_dashboard}); 
(C) A type of Growth Chart displaying the total number in the past week against the total number over time (by Aatish Bhatia~\cite{aatishb_growth_chart}); 
(D) Days since confirmed cases first reached 30 cases per day using event alignment (in a log scale) (by Our World In Data~\cite{owidcoronavirus}).}
\label{fig:temporal}
\end{figure*}

\textbf{(5) Communicating risk, vulnerability, and equity:}
Another set of visualizations aimed to communicate risk and unpack issues of vulnerability and equity (124 out of 668, 19\%). Previously identified potential risk factors for COVID-19 morbidity and mortality include age, race/ethnicity, gender, some medical conditions, the use of certain medications, poverty, crowding, certain occupations, and pregnancy~\cite{cdc_risk_factor}. These factors are also influenced by the social determinants of health, that is, social and economic conditions that influence the health of individuals and communities, such as income, education, employment, social support, and access to health care~\cite{lowcock2012social}.   

\textbf{(6) Gauging the multifaceted impacts of the crisis:}
Several visualizations in our corpus depicted the initial impacts of the crisis (141 out of 668, 21\%), including \textit{government response and interventions}; \textit{economic disruptions} such as changes in unemployment, GDP, S\&P ratings, and business sales; \textit{social disruptions} due to school closures, quarantine, and lockdowns in response to government policies; and \textit{environmental impact} such as the change of urban pollution and vibration of the Earth's surface. The impact of the pandemic has been multifaceted and wide-ranging. Visualizations in this category attempt to convey the shifts in daily life due to the crisis, and also serve as evidence of the effectiveness of interventions designed to mitigate risks. 

In summary, COVID-19 visualizations have been created to communicate a wide variety of messages. This message categorization provides an overall understanding of the focal areas of the visualizations in our corpus. In the next section, we use this message categorization to guide our in-depth examination of the visual approaches used to convey various messages about the pandemic.

\section{In what visual forms was the information presented?}
\label{sec:visual_forms}

We now turn to a discussion of how various visualization techniques and visual encodings have been used to create COVID-19 crisis visualizations that communicate each type of message identified above, along with some visualization examples. We also identify issues and challenges in existing COVID-19 visualizations, which will catalyze future research that tackles the challenges inherent within designing visualizations that address each type of messaging goal. Note that some screenshots used in this paper were cropped, and some visualization captions were not fully displayed here due to the space consideration. Care is needed when interpreting these visualization examples (e.g., to avoid potentially misleading messages or biases).
  
\subsection{Informing of the Severity}
\label{sec:inform_severity}

\begin{figure*}[tb]
\centering
\includegraphics[width=\textwidth]{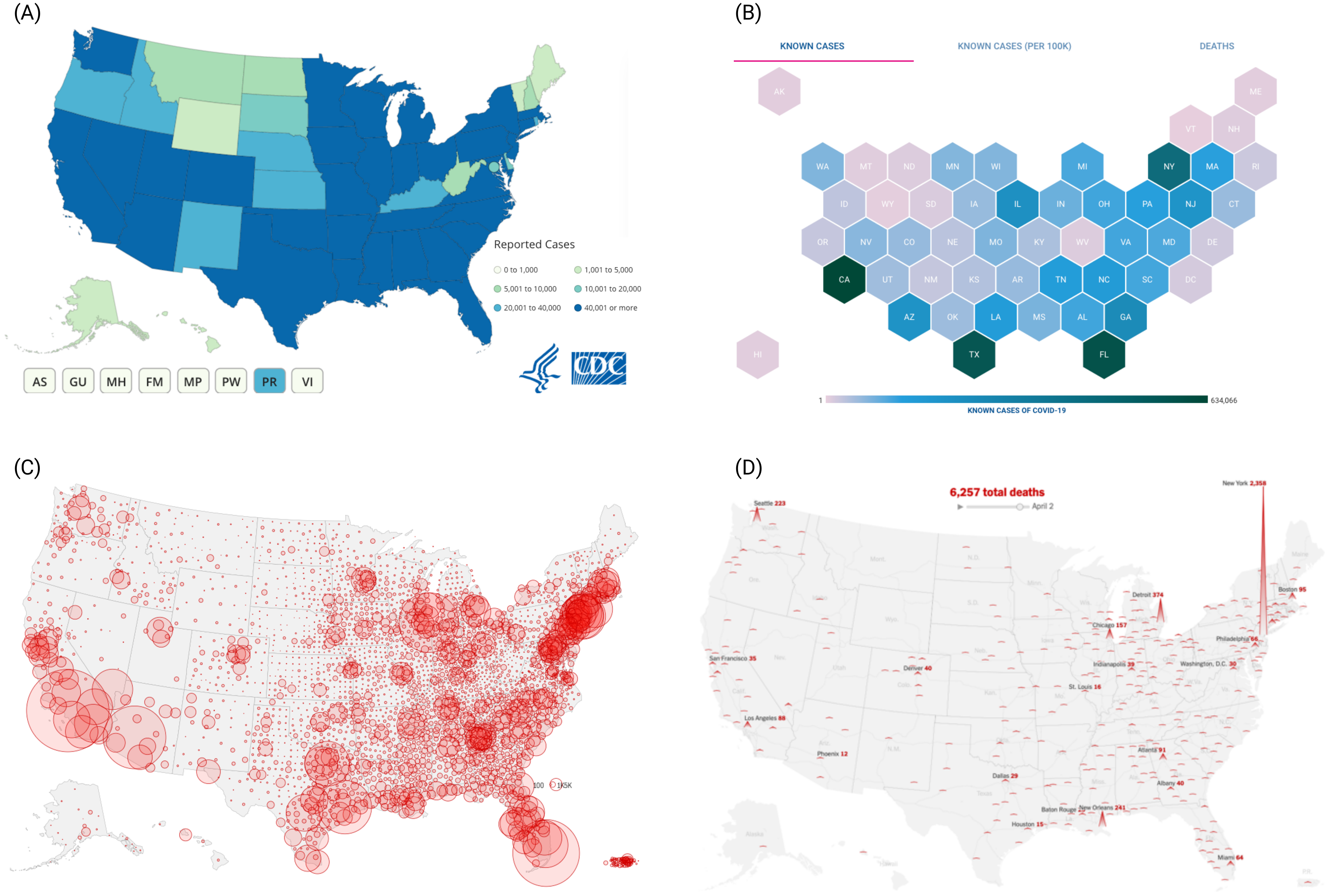}
\caption{Examples of geospatial visualizations: 
(A) A choropleth map (not normalized) using a traditional projection plots the total number of cases (by the Centers for Disease Control and Prevention~\cite{cdc_choropleth_map});
(B) A choropleth map overlaying on a tile grid base map provides both the raw number version and a normalized view (by the USAFacts~\cite{usafacts_tilegrid_map}); 
(C) A proportional symbol map (bubble map) shows the total cases (by CNN~\cite{cnn_bubble_map});
(D) An alternative proportional symbol map shows the total deaths over time. Each triangle represents the number of deaths in a metropolitan area (by the New York Times~\cite{nyt_triangle_map}).  
}
\label{fig:geospatial}
\end{figure*}

Visualizations that depicted the intensity of the pandemic mainly include traditional basic chart-based temporal visualization and map-based geospatial visualizations, and more advanced multivariate visualizations, as well as data-driven storytelling approaches. 

\textbf{Temporal visualizations} focused on depicting the trajectory of the pandemic over time (see \autoref{fig:temporal}). The most basic solution was to visualize daily and cumulative numbers over a time period, mostly using bar charts, line charts, and area charts. Two major variations in visualizing the temporal change of the pandemic include using logarithmic (log) scales (e.g., \autoref{fig:temporal} (C-D)) and adding the moving average that minimizes random fluctuations in counts (e.g., \autoref{fig:temporal} (A)). By the time of our review, 14 visualizations provided both a log- and a linear-scale mode (10 enabled users to toggle between modes), 6 provided only log scales, and the rest used linear scales. The wide adoption of linear scale charts may be because they are easier to understand for the general public.

Novel visualization techniques were introduced, such as the Pez Charts~\includegraphics[height=\fontcharht\font`\B]{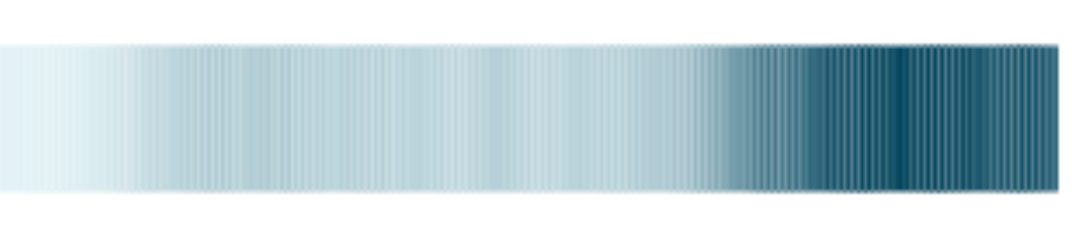} as seen in ~\autoref{fig:temporal} (B), the Growth Charts~\includegraphics[height=\fontcharht\font`\B]{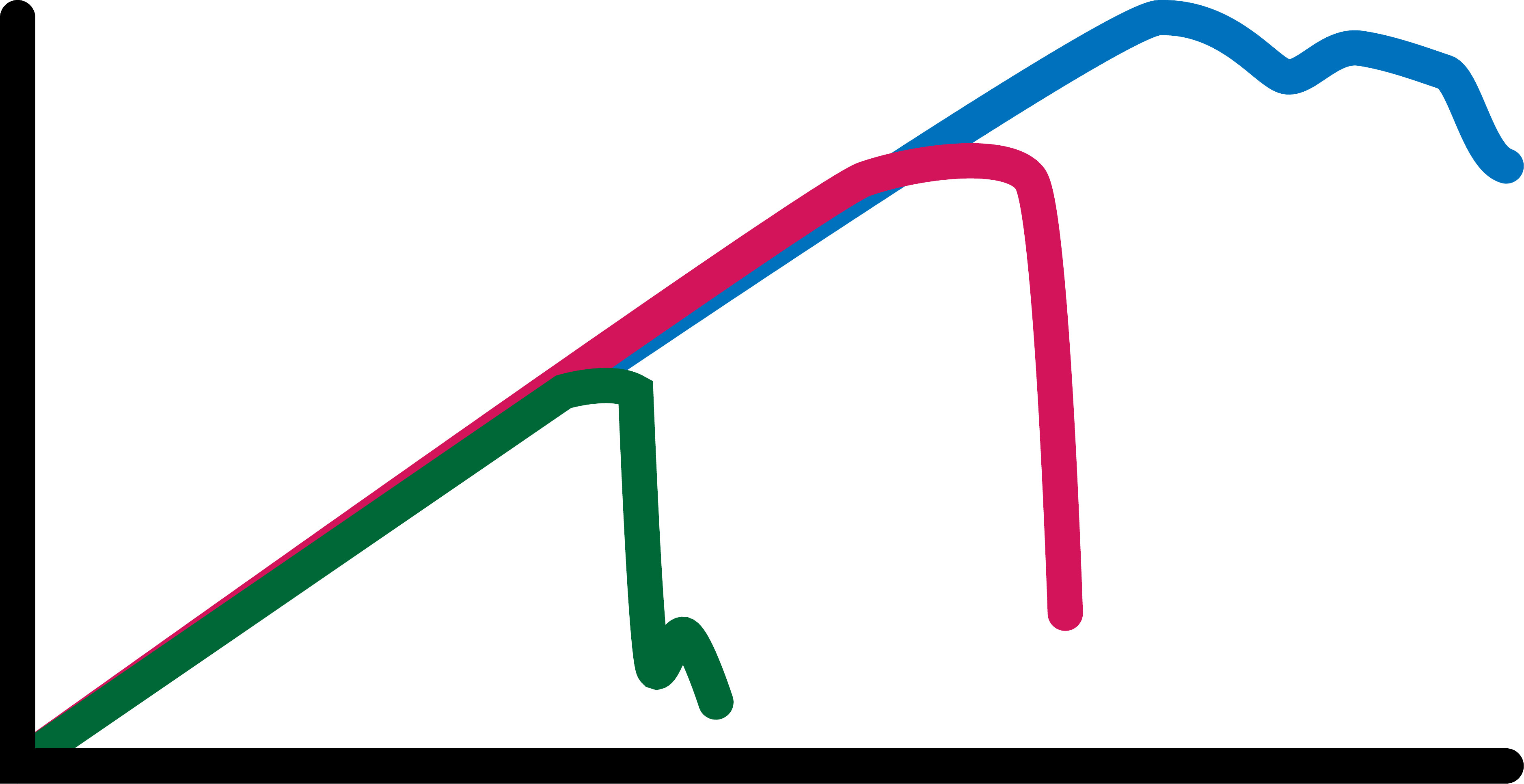} as shown in ~\autoref{fig:temporal} (C), and the use of temporal event alignment. 
The Pez Chart~\includegraphics[height=\fontcharht\font`\B]{fig/pez_chart_new.pdf}~\cite{pez_charts} aims to show case number changes in horizontally-placed blocks, with the x-axis representing time and the color of each block representing the values. When stacking multiple horizontal series together, it is easier to compare trends across regions. The first visualization that used the Pez Chart to show trends of the pandemic, in our corpus, was from the Los Angeles Times on April 1, 2020.
Moreover, \textit{temporal event alignment} is useful to reveal patterns that emerge by aligning the occurrence of events of interest over time~\cite{zhang2019eva}. The trajectory of the pandemic can be seen through an alignment of days because cases or deaths across regions (on the x-axis) are usually presented in a log scale. Comparing with the reference lines (i.e., the slope of a curve) helps demonstrate a doubling rate.  

Another emerging solution is the \href{https://aatishb.com/covidtrends/}{Growth Chart}~\cite{aatishb_growth_chart} \includegraphics[height=\fontcharht\font`\B]{fig/trend_chart.pdf} (see ~\autoref{fig:temporal} (C)), which plots the total number of the confirmed cases as the x-axis and the weekly confirmed cases as the y-axis~\cite{aatishb_growth_chart}. Applying the log scale to both axes and using the total case number as the x-axis (rather than time) helps reveal trends and patterns (e.g., demonstrating if cases are exponentially growing in a region or if a region is on the path to containing the virus).

\begin{figure*}[tb]
\centering
\includegraphics[width=\textwidth]{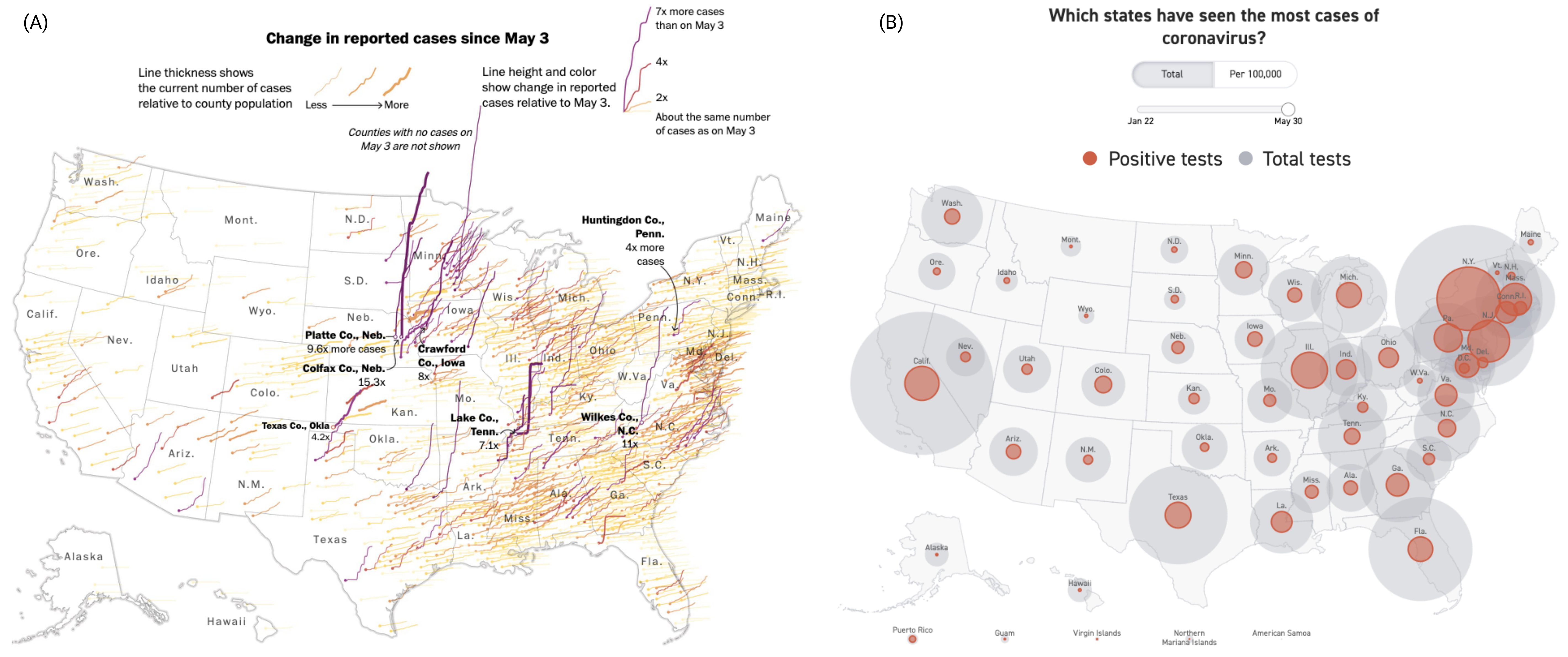}
\caption{Examples of multivariate visualizations: 
(A) Multiple encodings to illustrate reported case change over time, including using the line thickness to show the current number relative to population and using line height and color to show the change in reported cases relative to a specific date (as a type of event alignment) (by the Washington Post~\cite{wp_sparkline_map}); 
(B) Concentric/ centered bubble maps showing both the positive and total tests (by POLITICO~\cite{politico_centricbubble}). }
\label{fig:multivariate}
\end{figure*}

Though these more advanced methods have been studied and used before in academic research, they were not initially used for communicating information about the pandemic with the general public. The patterns conveyed in these novel visualizations are particularly relevant.

\textbf{Geospatial visualizations} displayed variables of interest over geographical maps to demonstrate which regions were impacted and compare how the impact differs by regions. Among all the map-based visualizations we analyzed, choropleth maps were the most popular option (110 out of 218, 50\%), followed by proportional symbol maps (61 out of 218, 29\% bubble maps). ~\autoref{fig:geospatial} shows four types of representative maps.

Both choropleth and proportional symbol maps overlay information on a base map (i.e., maps containing reference information that may use different geospatial information depending on what the designer wants to communicate~\cite{dent2008thematic}). Most map-based visualizations in our collection used maps that represented the physical shapes of geographical regions. Traditional projections (e.g., Mercator or Robinson) are widely used, while a few others in our collection used the Cahill–Keyes projection  ~\includegraphics[height=1.4\fontcharht\font`\B]{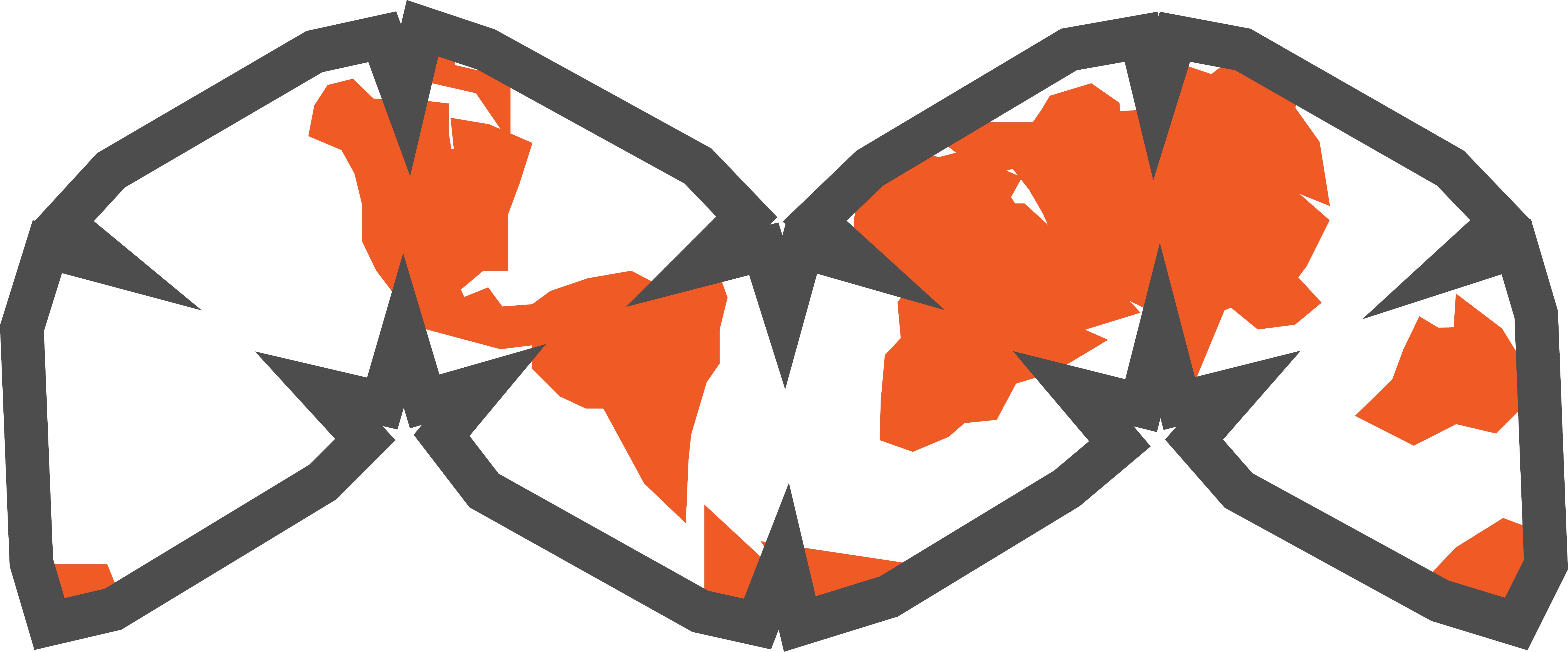}~\cite{nyt_cahill_projection}, and the Armadillo projection \includegraphics[height=1.4\fontcharht\font`\B]{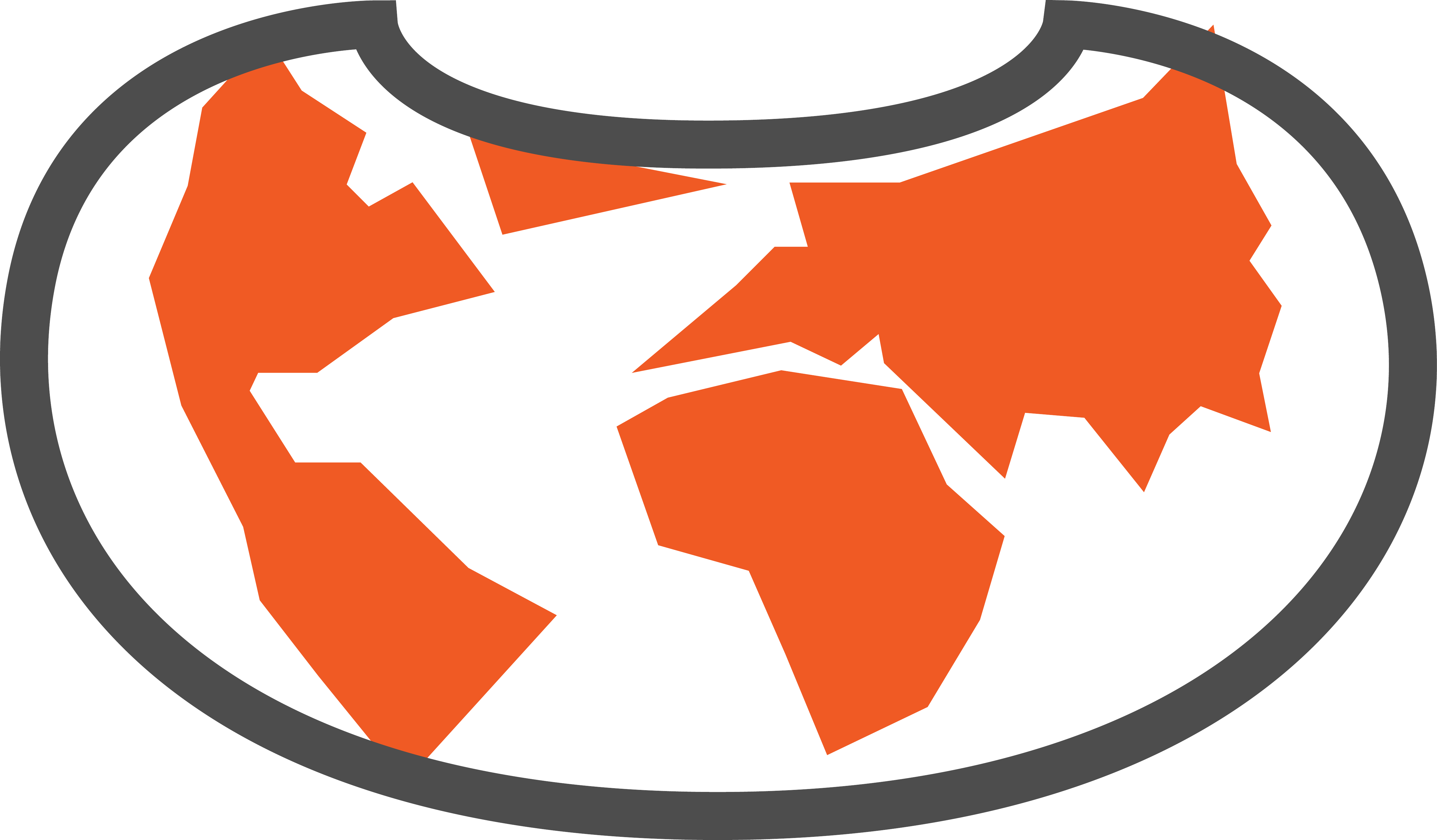} to provide a perspective with less distortions~\cite{scmp_armadillo_projection}. Different projections distort the globe in different ways, potentially impacting how the message is received by the audience. It was also common for visualizations to use tile grid maps, abstracting away the physical shape of geographical regions so that viewers can focus on the overlaying information. 


While choropleth maps were the most frequently used geospatial visualization in our corpus, they do present a challenge in that color-coding raw data may mislead viewers. For example, if a highly-populated state has the same case number as a less-populated state, the color will be the same on the map. A reader may interpret the map as conveying that the intensity of cases is similar in the two states, while in actuality, the intensity in the less-populated state is higher. Prior work has shown that normalizing data may help remove this type of bias~\cite{dent2008thematic}. Typical normalization variations include normalizing by area, relevant population, a prior date, central tendency (e.g., mean, median, mode), and variability (e.g., standard deviation, above and below range)~\cite{foster2019map}. However, our findings suggest that only 37\% of the visualizations in our corpus used some form of normalization. 


In addition, maps have applied various color schemes, including diverging (28\%), sequential (65\%), and a mix of diverging and sequential methods (7\%). Moreover, a majority of maps applied light background (e.g., white, light grey), but 17\% used dark background (e.g., black, dark blue). Prior work has examined color-related biases (e.g., dark-is-more, contrast-is-more bias) and the choice of background color~\cite{schloss2018mapping}. These design choices, while nuanced, may also impact how people perceive severity and risk in times of crisis.

\begin{figure*}[tb]
\centering
\includegraphics[width=\textwidth]{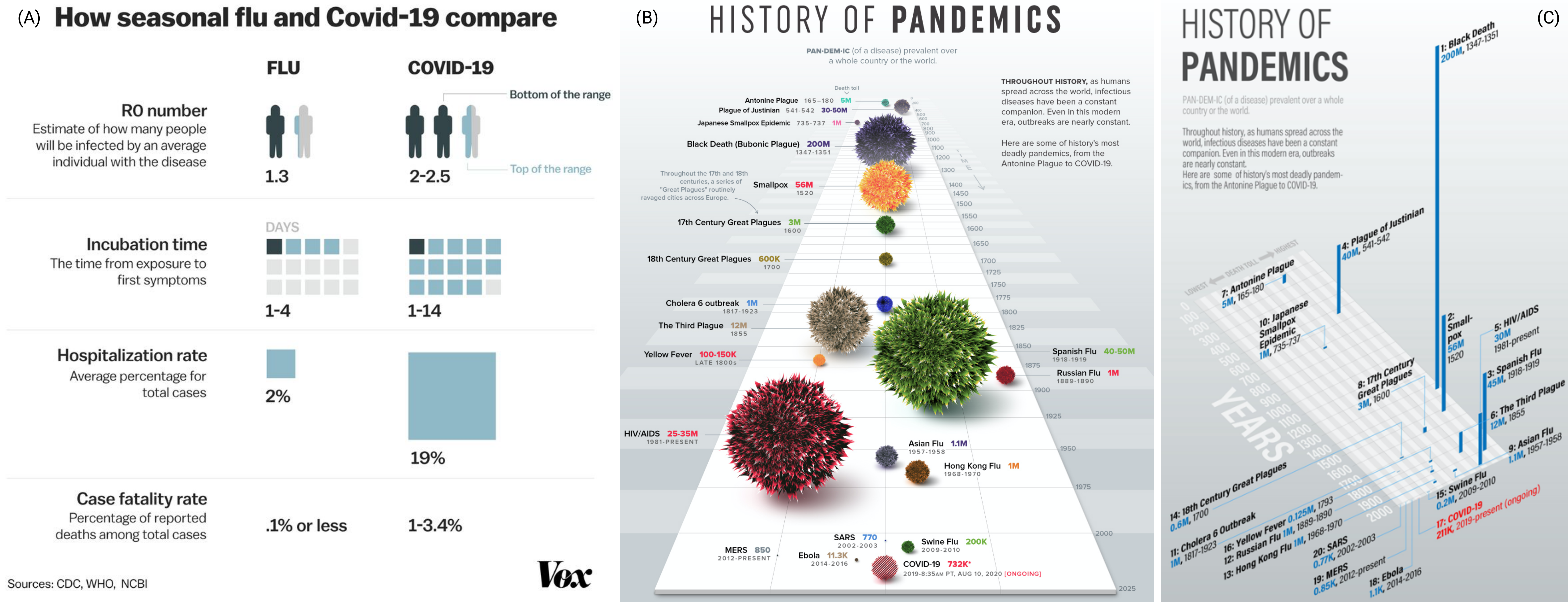}
\caption{Example visualizations of comparing with historical events: 
(A) A side-by-side table-like visualization that compares a variety of dimensions between the flu and COVID-19 (by Vox~\cite{vox_compare_history}); 
(B) A 3D view visualization that uses horizontal timelines along the Z-axis to present the history of pandemics (by Nicholas LePan~\cite{vc_hairball}); 
(C) A redesigned ``distorted'' 3D version of (B) shows the death toll of historical pandemics (by Ryu Sakai~\cite{medium_ryu}).}
\label{fig:history_compare}
\end{figure*}

\textbf{Multivariate visualizations} display two or more variables in a visualization, that aim to allow people to distill a complex multidimensional dataset and explore the patterns between variables. 
In our corpus, 8 visualizations used multivariate approaches, such as superimposing small multiples on maps to show cumulative number of cases over time in different locations (e.g., sparkline maps), as shown in ~\autoref{fig:multivariate} (A), using animation to illustrate how things have changed over time, as shown in ~\autoref{fig:multivariate} (B), and using concentric bubble maps to show multiple metrics (e.g., positive and total tests) in absolute numbers. Though it is common to use multivariate visualizations in visualization research and public health data analytics among professionals, it was not commonly used in our corpus. 

Visualizations that aim to provide \textbf{comparisons with historical events} mostly used table-like representations (\autoref{fig:history_compare} (A)) to compare characteristics between COVID-19 and other pandemics. Other visualizations embed 3D timelines~\cite{vc_hairball, medium_ryu} for easy comparisons, as shown in \autoref{fig:history_compare} (B-C).

\begin{figure}[tb]
\centering
\includegraphics[width=\linewidth]{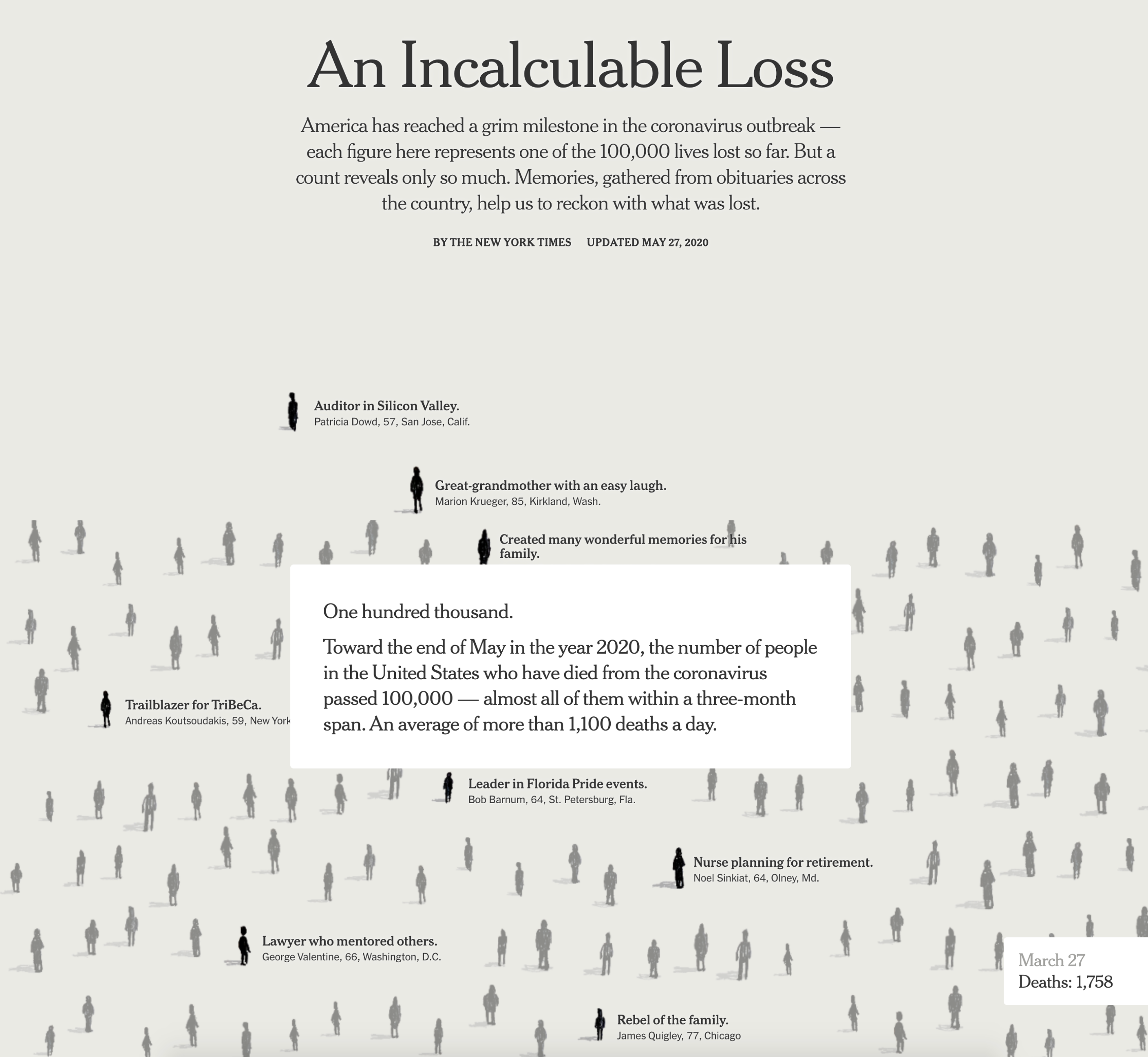}
\caption{Examples of acknowledging grief and sadness due to the loss of life: an interactive visualization using pictorial forms and scrollytelling, a visual storytelling approach that happens when visualization is revealed or changed as a user scrolls (by the New York Times~\cite{nyt_loss_interactive}). }
\label{fig:griefloss}
\end{figure}

\textbf{Crisis visualizations that \textit{turned to the felt experience}} were also designed to convey the severity of the pandemic, though unlike the previously-described data visualizations. For example, the visualizations created by the New York Times (\autoref{fig:griefloss}) had focused on personalization and emotions. These visualizations convey a sense of acknowledging grief and sadness resulting from the loss of lives due to the pandemic. Though we only found two visualizations in this category, they carried an important message that the death toll was not only a number---it reflects an indescribable loss of valued human life. 

\textbf{Summary:} Our findings show various trends in the use of visual encodings, traditional chart-based techniques, and novel data visualization approaches within our corpus of COVID-19 visualizations. We identified some common issues and challenges within these visualizations, such as the lack of normalization. We also presented data-driven storytelling approaches to convey the severity of the pandemic. Future work should further examine the effect of different types of visualizations on people's risk perception, and identify the most and least effective approaches for visualizations.

\subsection{Forecasting Trends and Influences}
\label{subsec:forecasting}

\begin{figure*}[tb]
\centering
\includegraphics[width=\textwidth]{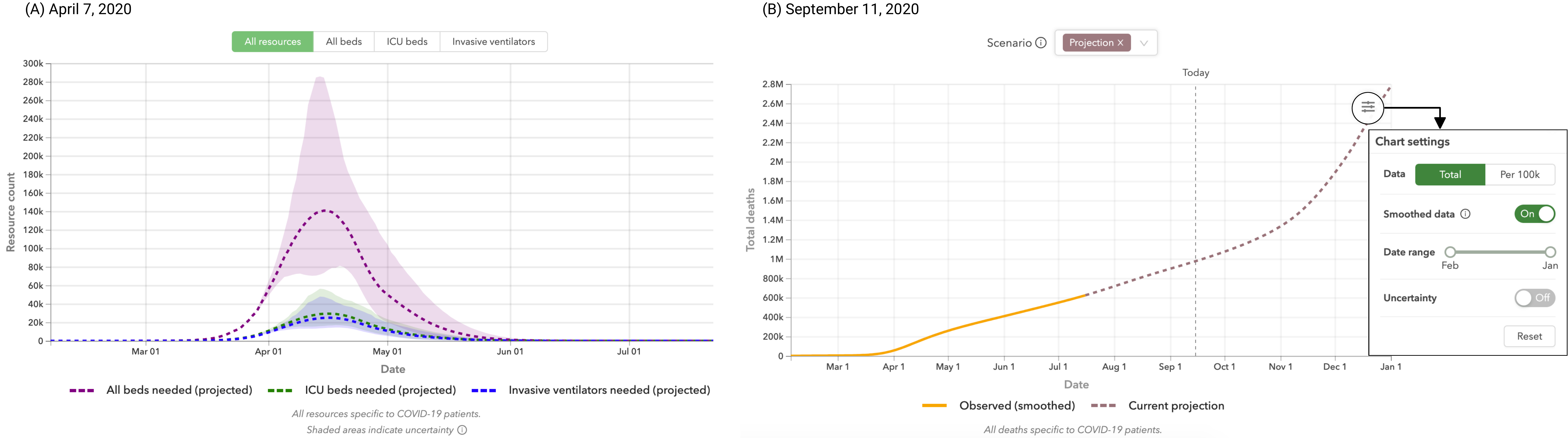} 
\caption{Examples of the uncertainty visualizations using different visual encodings at different time (by the University of Washington~\cite{uw_model}). (A) A visualization with uncertainty annotations with shades (April 7, 2020); (B) A visualization without uncertainty annotations (no shade) as the default view (September 11, 2020).} 
\label{fig:uncertainty_switch}
\end{figure*} 

\begin{figure*}[tb]
\centering
\includegraphics[width=\textwidth]{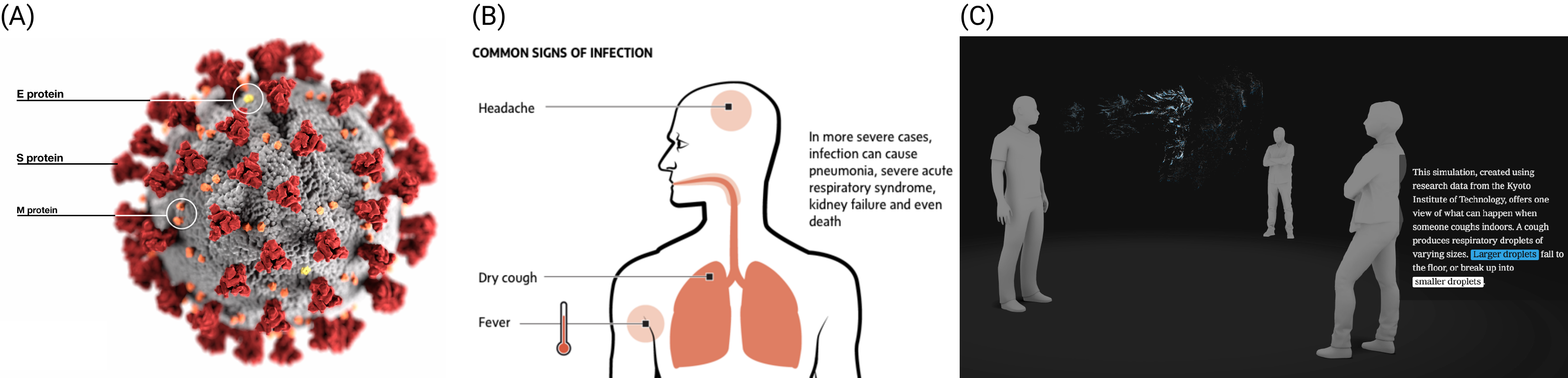}
\caption{Example visualizations of explaining the nature of the crisis: 
(A) A medical illustration created at the Centers for Disease Control and Prevention (CDC) reveals ultrastructural morphology exhibited by coronaviruses (by the CDC~\cite{cdc_coronavirus_illustration}); 
(B) A type of symptom-body map shows associated symptoms on the body shape (by The Globe and Mail~\cite{globemail_bodymap});
(C) A 3D visualization using scrollytelling vividly describes how possible transmission happens (by the New York Times~\cite{nyt_3d_simulation}).
}
\label{fig:course_of_diseaase}
\end{figure*}

Forecasting models intrinsically involve a certain level of uncertainty. Many visualizations in this category use graphical annotations to quantify uncertainty (48 out of 57, 84\%). Among the collected visualizations for forecasting and projection, major approaches to communicate uncertainty include: presenting prediction intervals (29 out of 48, 60\%), comparing projection results of multiple models or different scenarios (14 out of 48, 29\%), and communicating uncertainty in text (5 out of 48, 10\%). 
Visualizing uncertainty intervals for projections (n=29) includes using solid line charts with shade \includegraphics[height=1.2\fontcharht\font`\B]{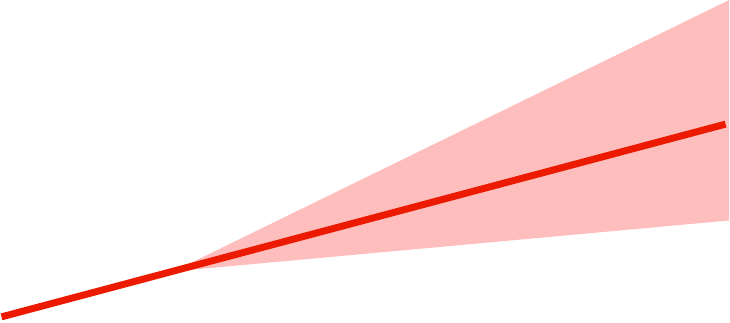} (16 out of 29, 55\%), a solid line indicating current and prior situations followed by a dashed line with shade indicating projection \includegraphics[height=1.2\fontcharht\font`\B]{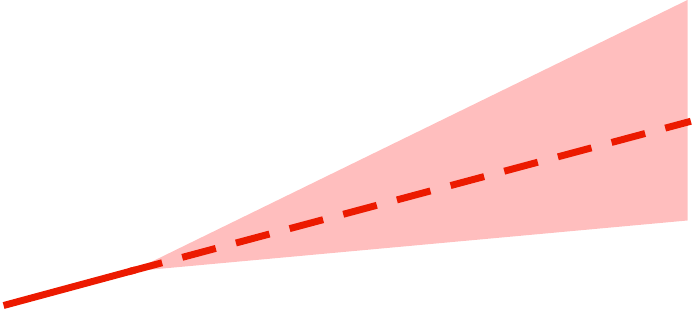} (6 out of 29, 21\% ), superimposing multiple levels of confidence intervals (4 out of 29, 14\%) shown with different opacity using interval funnels \includegraphics[height=1.2\fontcharht\font`\B]{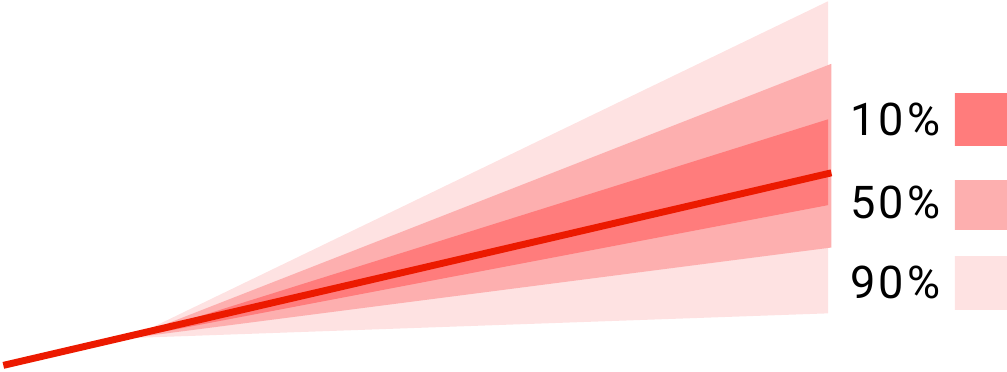} and interval range bars  \includegraphics[height=1\fontcharht\font`\B]{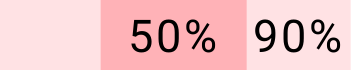}, as well as showing confidence intervals with error bars (3 out of 29, 10\% for visualizing effective reproduction rate). 

As more and more forecasting models are generated, visualizations are more commonly used to compare between different models or scenarios, e.g., using ensemble plots (n=9) and multiple hypothetical outcomes (n=5). The goal is to present differences in the assumptions and corresponding confidence intervals and to help the public understand the potential limitations of the forecasts. Multiple hypothetical outcomes can be useful for understanding various scenarios regarding what might happen in the future. For example, in our corpus, visualizations that juxtapose small multiple choropleth maps in the same view~\includegraphics[height=1.2\fontcharht\font`\B]{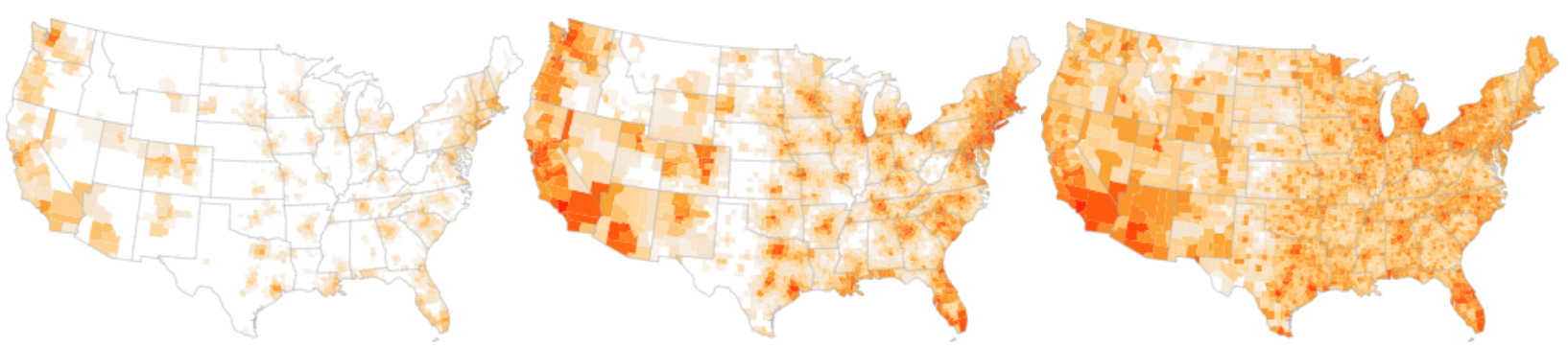} show how different infection prevention and control measures might influence the outbreak's spread~\cite{nyt_sm_control}. This type of visualization may be able to intuitively help people understand the probability amidst this complex pandemic. 

Among all the forecasting visualizations in our corpus, we noticed one case where the default view of the visualization switched from showing uncertainty annotations with shade (\autoref{fig:uncertainty_switch} (A)) to charts without uncertainty annotations (\autoref{fig:uncertainty_switch} (B)). Instead, users were able to toggle on and off the button \includegraphics[height=1\fontcharht\font`\B]{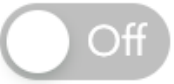}.

\textbf{Summary:} Our findings summarize the design choices of whether to present uncertainties and the visual styles to communicate uncertainty among the COVID-19 visualizations. Future research needs to examine how to effectively communicate uncertainty to lay audiences to increase credibility and improve decision-making. Understanding the nuanced differences in visual encodings for uncertainty will help contribute to the increasing body of literature in uncertainty research.

\subsection{Explaining the Nature of the Crisis}
\label{sec:explain_course_of_disease}

\begin{figure*}[tb]
\centering
\includegraphics[width=0.95\textwidth]{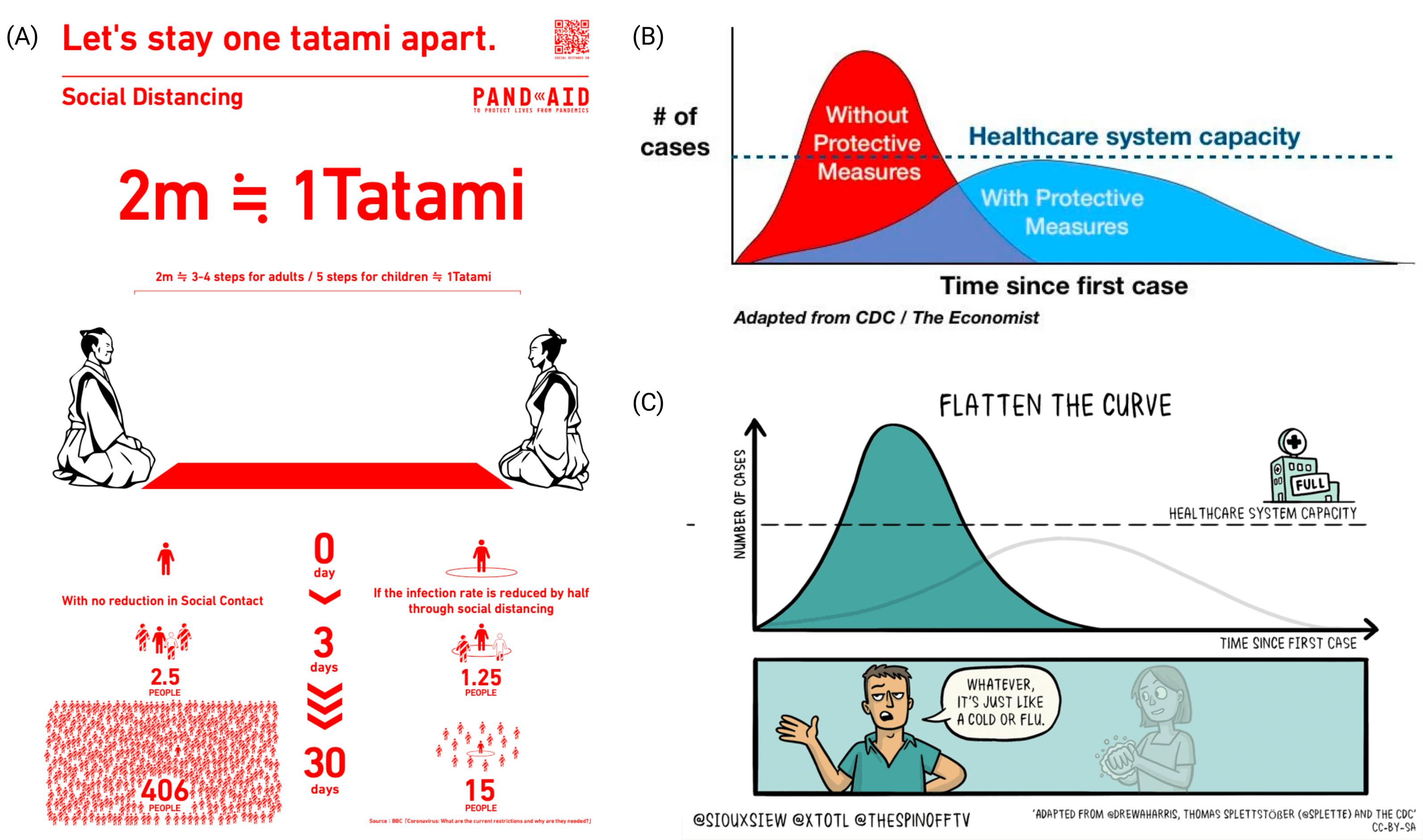}
\caption{Example visualizations focused on guiding risk mitigation: 
(A) An infographic designed by a group of volunteers in Japan using the cultural metaphor of tatami, to show how to engage in social distancing (by PANDAID~\cite{jp_tatami}); 
(B) A flatten-the-curve illustration (by Drew A. Harris~\cite{harris_flatten_twitter});  
(C) An animated cartoon version of the flatten-the-curve illustration (by Dr Siouxsie Wiles et al.~\cite{siouxsie_flatten_animated}).}
\label{fig:risk_mitigation}
\vspace{-10pt}
\end{figure*}

Visualizations that \textbf{explained the causes, symptoms, and transmission} were mainly designed to facilitate communication of information generated by health professionals and scientists. For example, the medical illustrations team from the Centers for Disease Control and Prevention (CDC) aimed to make abstract medical concepts more approachable and comprehensible, as shown in~\autoref{fig:course_of_diseaase} (A):
\textit{``The colours were chosen for visual impact. The bold red of the S proteins contrasted by the gray of the viral wall, adds a feeling of alarm... Shadows add to the realism''}~\cite{cdc_coronavirus_illustration}. The choices of colors and shadows may be meant to spark the felt experience (i.e., how visualizations are felt by the users). 


Besides, visualizations have also been created to help people \textbf{understand and compare symptoms} between COVID-19 and other conditions (e.g.,  allergies, cold, and, influenza). Side-by-side table-like visualizations have been widely used for symptom comparison. Some characteristics of this type of visualizations include using categorical color encoding to represent various frequencies of symptoms (like rare, sometimes, and common), and embedding pictorial aids next to the text. While side-by-side tables are easy for comparison, \textit{symptom-body maps} that map associated symptoms on the body shape (e.g.,~\autoref{fig:course_of_diseaase} (B)) help make the invisible become visible. 

Messages that aim to \textbf{explain transmission} focus on explaining how the virus spreads or the various ways in which it can spread. For example, a 3D simulation visualization, as shown in ~\autoref{fig:course_of_diseaase} (C)~\cite{nyt_3d_simulation}, vividly explains the possible transmission routes and attempts to persuade people to keep social distance. To that end, these visualizations also aim to help people \textit{understand the reasonableness} of recommended actions (e.g., social distancing) and reinforce personal responsibility to reduce harm to other individuals and the community~\cite{reynolds2007cdc}. 

\textbf{Visualizations of contact tracing} can be classified into explaining the process behind contact tracing and visualizing individual-level or group-level tracing. At an individual level, visualizations illustrate patient movement paths and timelines \includegraphics[height=1.2\fontcharht\font`\B]{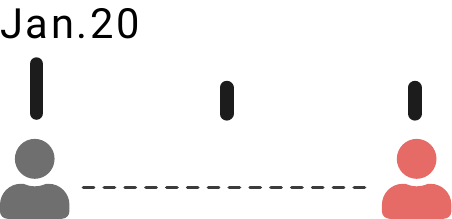} to explain how an individual might have come into contact with a person infected with the virus. At the group-level, node-link networks~\includegraphics[height=1.2\fontcharht\font`\B]{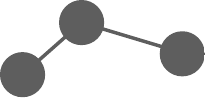} explore clusters of core groups or super spreaders.  

\textbf{Summary:} Communicating scientific research to the general public is important for scientists, yet it is also challenging to effectively communicate with diverse audiences~\cite{brownell2013science}. Visualization designers have made efforts to help bring science-based information closer to the public. Prior work suggests using pictorial aids in health communication is an effective way to facilitate understanding and improve recall of medical instructions~\cite{arcia2019helping}. Though it is promising to incorporate pictorial forms to help interpret textual explanations, it is important to keep in mind that these visual aids might be too complex to understand or they may fail in reflecting viewers' expectations~\cite{katz2006use}.    


\subsection{Guiding Risk Mitigation}
\label{sec:guiding_risk_mitigation}

\textbf{Providing instructional guidance} was one of the common approaches to guiding the public in risk mitigation. Infographics, especially instructional graphics, have been produced and distributed widely to provide instructional guidance on mitigating strategies, such as how to maintain social distancing, use personal protective equipment~(PPE), and maintain hygiene. 
Several visualizations (n=6) have incorporated some sort of cultural metaphors in the visualizations that aim to guide people on how to avoid risks. 
For example, social distancing guidance designed by a Japanese organization adopted the metaphor of tatami (a type of mat in traditional Japanese-style rooms) to guide people to ``stay one tatami apart''~\cite{jp_tatami} (see ~\autoref{fig:risk_mitigation} (A)), while a Florida county reminded people to ``stay one alligator apart''~\cite{alligator_social_distancing}.

\begin{figure}[tb]
\centering
\includegraphics[width=\linewidth]{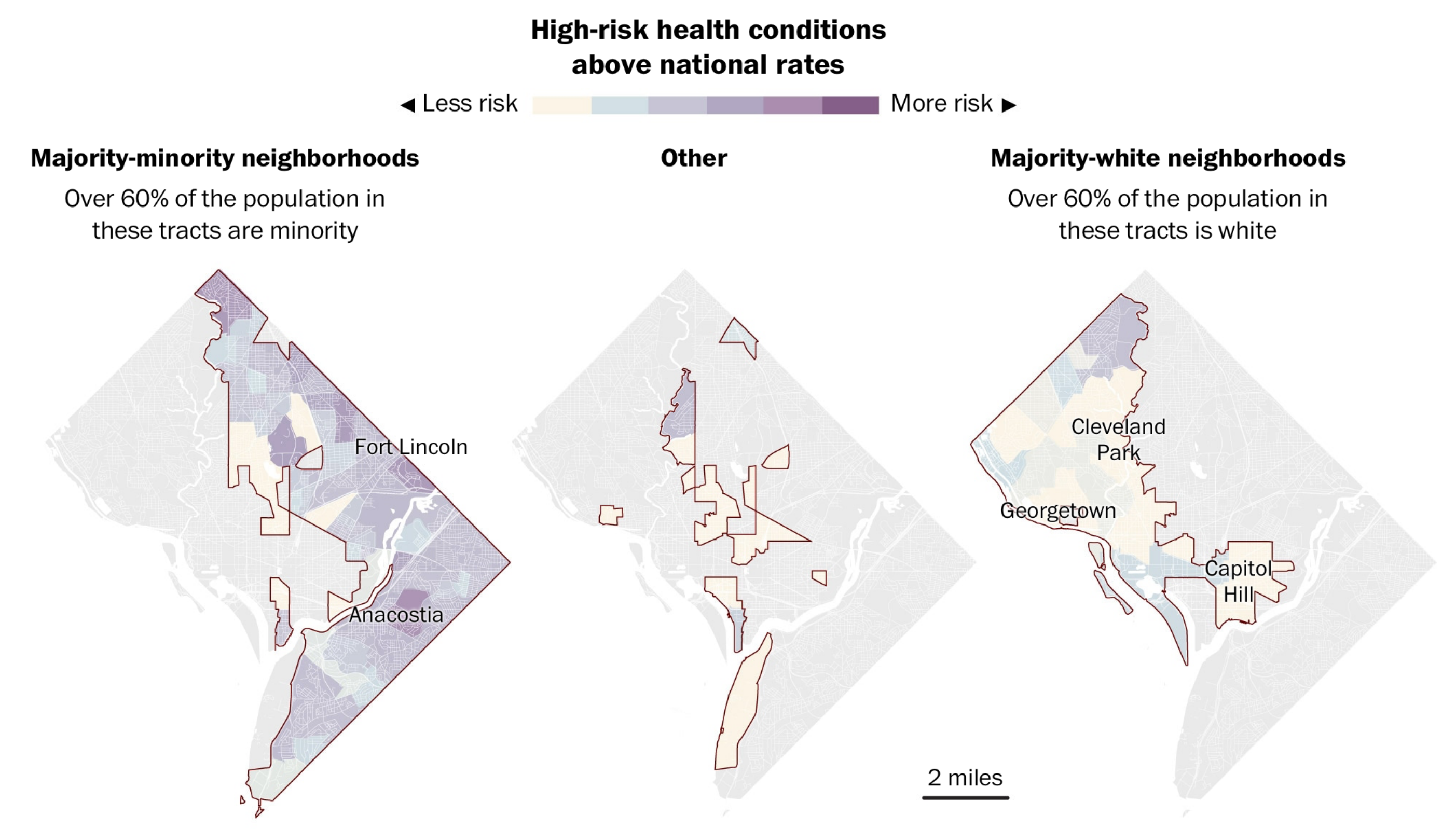}
\caption{A set of choropleth maps show how COVID-19 further demonstrates existing health disparities in communities of color (by the Washington Post~\cite{wp_health_disparities}). }
\label{fig:wp_health_disparities}
\end{figure}

\begin{figure*}[tb]
\centering
\includegraphics[width=\textwidth]{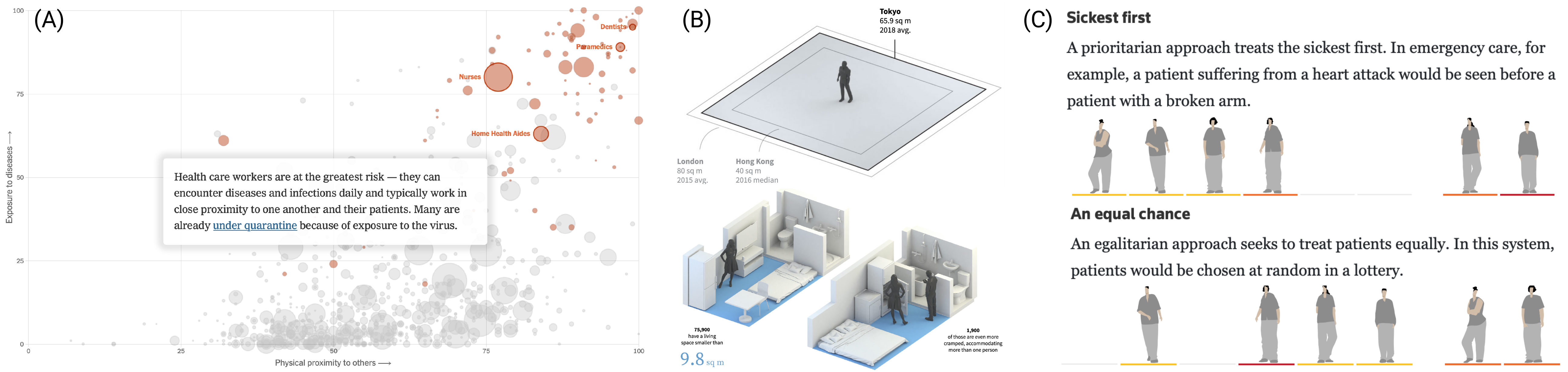}
\caption{Examples of visualizing risk, vulnerability, and equity: 
(A) Scatter plots using scrollytelling show people who face the greatest risk of exposure to coronavirus according to the physical proximity to others (by Lazaro Gamio~\cite{nyt_vulnerable_occupation}); 
(B) 3D visualizations vividly depict the challenge of sheltering in limited spaces (by Reuters~\cite{reuters_small_shelters});
(C) A scrollytelling visualization explains approaches to rationing care, asking the question \textit{``when medical resources are limited, who should get care first?''} (by Feilding Cage~\cite{reuters_rationing_care}).  
}
\label{fig:equity}
\end{figure*}

\textbf{Conceptual flattening-the-curve} charts (e.g.,~\autoref{fig:risk_mitigation} (B-C)) have become prevalent, particularly with the addition of a horizontal line marking ``healthcare system capacity''. We found 21 different versions of flattening-the-curve charts that aim to conceptually depict the management of the healthcare systems' capacity. They appear to emphasize personal responsibility in minimizing the unprecedented strain on the health system. However, these type of visuals have become controversial amidst the pandemic as they simplify complex pandemic situations (e.g., implying that it is good enough to keep the patient counts below the healthcare system's capacity) and they may also be misinterpreted by the public.

\textbf{Summary:} Though we did not systematically analyze the cultural elements in our corpus of visualizations, it is important for future work to examine cultural differences in the design of crisis visualizations, and the effect of culturally-tailored visualizations on people's risk perceptions and behaviors. Moreover, more work is needed to examine the effectiveness of data-driven empirical visualizations as compared to conceptual visuals like the flatten-the-curve illustrations.

\subsection{Communicating Risk, Vulnerability, and Equity} 
\label{subsec:equity}

  
Our findings highlight a body of visualizations (7 out of 81, 9\%) that \textbf{used aggregated scores to communicate risks} (i.e., using a single value to reflect risks arising from multiple factors), such as \href{https://svi.cdc.gov/map.html}{the Social Vulnerability Index (SVI)} and Behavioral Risk Factor Surveillance System (BRFSS). The resulting visualizations were all choropleth maps (e.g., \autoref{fig:wp_health_disparities}), that revealed health disparities caused by multiple factors including majority-minority, highly vulnerable, overcrowded households, and most uninsured~\cite{wp_health_disparities}.

An increasing number of visualizations have added \textbf{disaggregated demographic data in their existing visualizations} as governmental authorities began to release demographic information of patients. Bar charts were used to display demographic distributions of affected populations. Age distribution was the most commonly visualized (n=39), followed by race (n=27) and gender (n=23). Some visualizations (n=5) unpacked the \textbf{influence of underlying health conditions} to understand the characteristics of affected population, and the likelihood of people seeking intensive care treatment.
Limited access to healthcare is another risk factor for severe disease~\cite{lowcock2012social}. People who have limited access to health care resources are likely to be more vulnerable during a public health crisis~\cite{reynolds2007cdc}. To help understand these factors, visualizations have been created to show \textbf{accessibility and allocation of resources} (n=4), such as visualizing the travel distance to nearest hospitals to get treatment and displaying the geospatial spread of tangible social support (e.g., food services).

In addition, visualizations have been created to \textbf{examine associations of socioeconomic status~(SES) and the pandemic} (15 out of 81, 19\%). Various types of visualizations have been adopted in this category, including scatter plots for displaying the relationship of physical proximity and the exposure to diseases, bar charts to show ranked order of jobs based on forced policies of physical distancing, and index charts that compare high and low income disparities in movement change.

Despite the effort to unpack and address the vulnerability and equity issues, we also found that visualizations in this category were not equally created across visualization outlets. In our corpus, visualizations created by the government agencies only show basic population distribution by age, race, and gender, with two visualizations displaying underlying condition. However, no government created visualizations examined other SES-related factors. Instead, work examining SES, accessibility to resources, and living conditions were created by the news outlets, independent media, companies, and NGOs, indicating the effort being put in revealing vulnerability and equity challenges, as shown in \autoref{fig:equity} (A-B). Scrollytelling visualizations have also been designed to explain who should get access to medical services first when resources are limited, e.g., sickest first, an equal chance, and maximizing treatment benefits~\cite{reuters_rationing_care}, as shown in \autoref{fig:equity} (C). Such visualizations highlight a socioeconomic challenge in terms of imperfectness of the healthcare system and may shed light on how to reform and improve the system.

\begin{figure*}[tb]
\centering
\includegraphics[width=\textwidth]{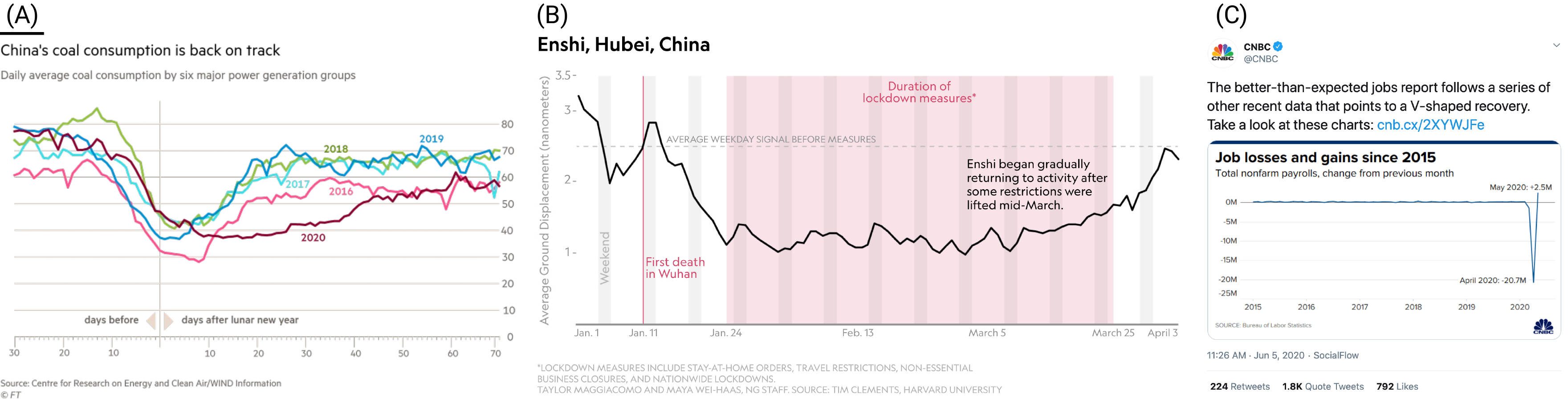}
\caption{Example visualizations of gauging the multifaceted impacts of the crisis:
(A) A superimposed line chart shows how COVID-19 stalled climate change momentum, using single event alignment technique with the vertical line indicating the alignment event of Chinese Lunar New Year (by the Financial Times~\cite{ft_lunar_alignment});
(B) A line chart visualizes how the level of the earth's surface vibration has changed over time during the lockdown periods. Shaded areas indicate the interval-event duration (by the National Geographic Society~\cite{ng_earthquite});
(C) A ``misleading'' V-shape-like line chart that went viral on Twitter (by CNBC~\cite{CNBC_tweet}).
}
\label{fig:impact_crisis}
\end{figure*}

\textbf{Summary:} Our findings show that a number of visualizations created by a variety of information outlets help communicate risks, and unpack issues around vulnerability and equity amidst the COVID-19 crisis. It is important to communicate that some populations face greater risks than others because it highlights the inequities that exist within society and calls attention to areas in which concerted efforts are needed to close gaps in terms of how different communities are impacted by a crisis. However, when visualizing risk, vulnerability, and equity, one caveat is that deliberate efforts need to be made to prevent and counteract stigmatization that can arise by focusing on how the pandemic is negatively impacting a group.

\subsection{Gauging the Multifaceted Impacts of the Crisis}
\label{sec:gauge_impact}

We identified two challenges for visualizations that aim to demonstrate the impacts of the pandemic on human lives and society.  
The first challenge is building the connection between the theme of the visualization and the major events amidst the crisis. Techniques that help building such connections include \textit{event alignment}, such as aligning by Lunar New Year to show the change of coal consumption in China, as shown in ~\autoref{fig:impact_crisis} (A), and event annotations that indicate when key events occurred, including \textit{point-event indicators without duration} (e.g., first death in Wuhan) and \textit{interval-event indicators} (e.g., duration shutdown order in effect), as shown in ~\autoref{fig:impact_crisis} (B).

The second challenge is how to vividly tell a story about the effect of certain interventions (e.g., public health and government interventions) without misleading audiences. One ``misleading'' visualization example that went viral on social media was the V-shape-like line chart presented by CNBC (see \autoref{fig:impact_crisis} (C)) to support the claim that the job market was recovered. The problem was the inappropriate choice of chart type. Many social media users pointed out this ``misleading'' chart issue in the thread of the discussion and proposed alternative design solutions like using various bar charts to improve the design.

\begin{figure*}[tb]
\centering
\includegraphics[width=\textwidth]{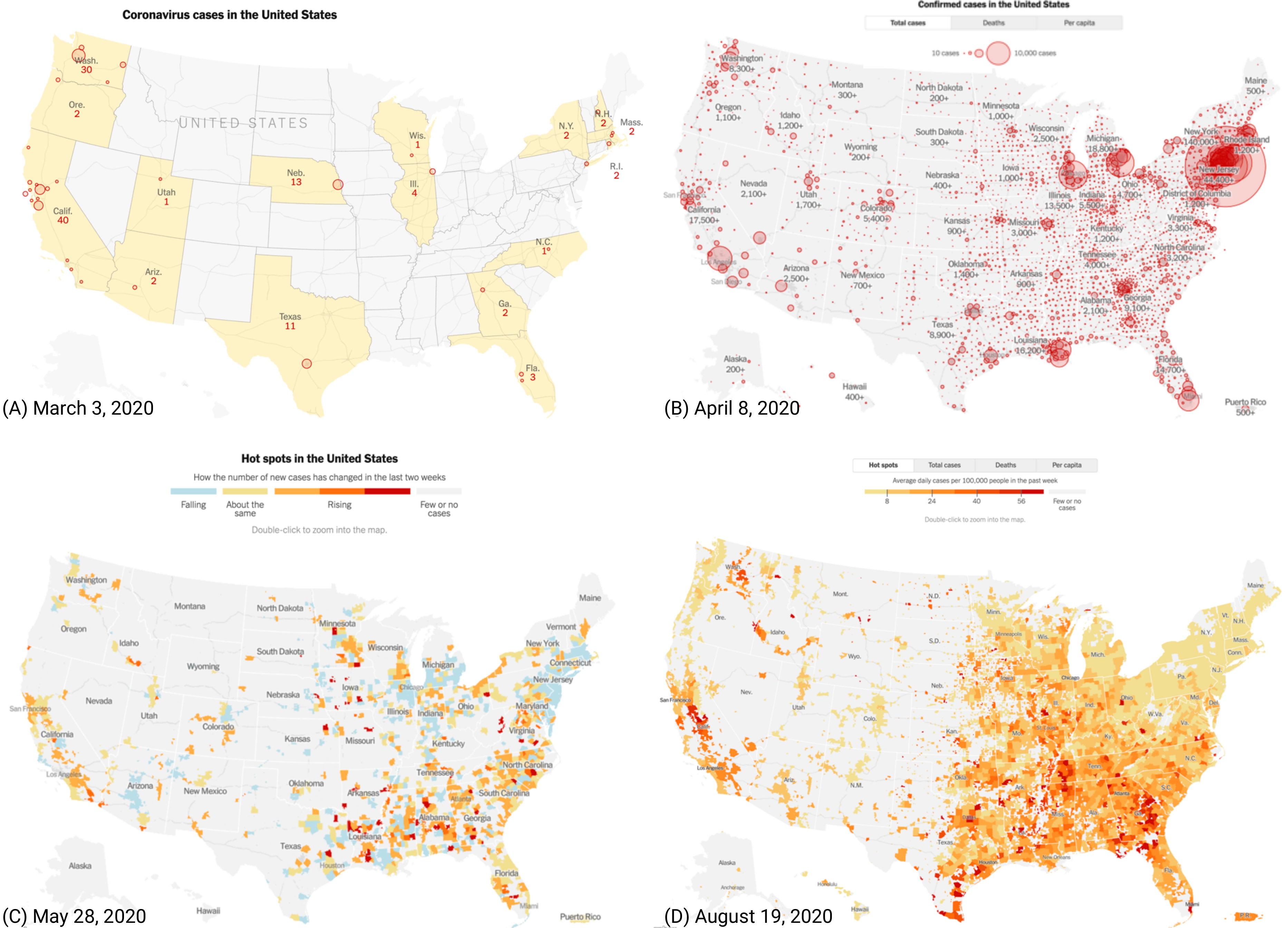}
\caption{Examples of how crisis visualizations have changed over time by the \textit{New York Times}~\cite{nyt_dashboard}. (A) A proportional symbol map (i.e., bubble map) superimposed on a qualitative map, published on March 3, 2020; (B) A bubble map published on April 8, 2020; (C) A choropleth map with categorical and sequential color schemes, published on May 28, 2020; (D) A choropleth map (with updated legend and menu), published on August 19, 2020.}
\label{fig:nyt_change_over_time}
\end{figure*}

\textbf{Summary:} Though the visualizations in this category seem to be general and can be designed at any time, we argue they are particularly important and relevant to crises. The design and dissemination of ``misleading'' visualizations in times of crisis pose extra challenges to the general public as these deceptive visualizations may impact people's trust of officials and organizations, and people's decision making. Considering the public engagement with existing crisis visualizations, we see how visualizations can be improved for more effective and responsible communication. 

\jz{ 
\subsection{Narrative Visualization Approaches among COVID-19 Visualizations}
One growing area of work in the visualization space is that of narrative visualization~\cite{Kosara2013Story}, that is, ``a genre that combines interaction techniques for exploratory control over insights gained and communicative, rhetorical, and
persuasive techniques for conveying an intended story''~\cite{hullman2011visualization, Lee2015More}. As a means of characterizing the growing landscape of narrative visualizations, Segel and Heer conducted a design space analysis that categorized visualization genres, visual narrative tactics, and narrative structure tactics~\cite{Segel2010Narrative}. Hullman and Diakopoulos's~\cite{hullman2011visualization} work on visualization rhetoric provided an analytical framework to help people understand how design cues facilitate prioritization of particular interpretations in visualizations. Moreover, Moere et al. \cite{moere2011role} argued for making the information visualization research more inclusive by incorporating more reflection and critiques from visualization practice work (e.g., visualization activities that are conducted by commercial enterprises and freelance designers). 
Though not all visualizations in our corpus are narrative visualizations ~\cite{Lee2015More}, some have applied narrative visualization approaches, such as annotation, animation, and scrollytelling. Building upon prior narrative visualization research, our work examined how narrative visualization approaches were applied in COVID-19 visualizations, with a focus on visualizations produced by practitioners (i.e., those available to the general public). } 

\jz{
In the preceding sections, we have summarized how various visualization forms have been used to communicate a range of messages in the COVID-19 pandemic (\autoref{sec:inform_severity} to \autoref{sec:gauge_impact}). Beyond the visual encodings and techniques discussed thus far, we also found that narrative visualization techniques were variably used to communicate COVID-19 messages. For example, animation was most often used (39\%) to depict the severity of the pandemic by showing how fast coronavirus has spread across countries and how current cases have changed over the course of the pandemic. Animation was also used in other contexts, although less frequently. Specifically, animation was used in visualizations designed to: explain virus transmission (17\%), unpack health equity and vulnerability (17\%), guide people towards mitigating risks (13\%), gauge the impact of the crisis (10\%), and forecast trends and influences (4\%). Interestingly we found that scrollytelling techniques were only present in visualizations created by news outlets (e.g., the Washington Post, the New York Times, and Reuters).
}

\begin{figure*}[tb]
\centering
\includegraphics[width=\textwidth]{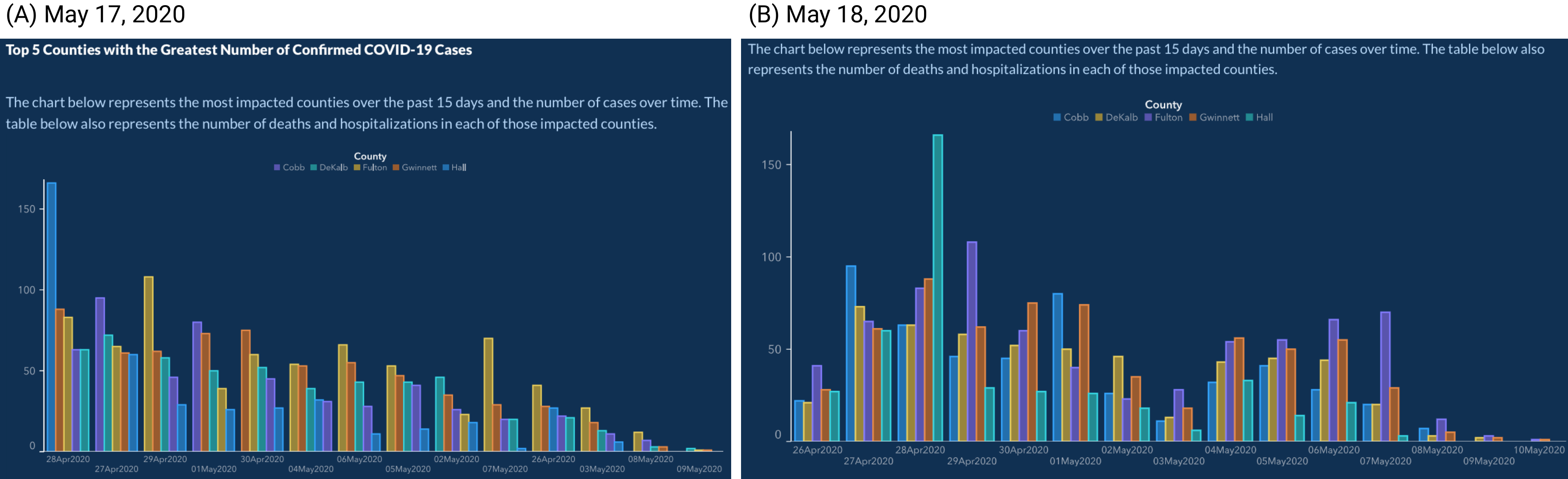}
\caption{Examples of a ``misleading'' visualization and its corrected version by the Georgia Department of Public Health~\cite{GA_misleading}.
(A) A bar chart aims to show that daily cases were declining in five of the state's hardest-hit counties in Georgia. But note that the dates were arranged out of order; 
(B) A corrected chart of (A) after being accused of data manipulation using misleading visualizations.}
\label{fig:GA_misleading}
\end{figure*}

\section{Under What Circumstances? Crisis Visualizations within Dynamic Temporal Contexts}
\label{sec:contexts}

So far, we have presented four components of our framework to examine COVID-19 visualizations. In this section, we draw attention to the complex and dynamic temporal contexts of the pandemic. Our findings show that 36\% of the visualizations in our corpus (243 out of 668) have been updated within our review time. 

We categorize these changes as being \textbf{crisis-driven}, \textbf{public-driven}, and \textbf{content producer-driven}. These categories are not likely to be mutually exclusive nor do they possess strict boundaries, and they are inherently subjective given that we are not privy to the intentions designers held as they were creating the visualizations. Indeed, in our discussion below, we do not report on the \textit{intentions} of visualization creators as these visualizations were changed, but rather the ways in which the changing visualizations in our corpus \textit{reflect} the dynamic nature of: the pandemic (``crisis-driven changes''), public response to created visualizations (``public-driven changes''), and the strategies used to create visualizations (``content producer-driven changes''). The benefits of categorizing such changes over time, are that it helps us understand how temporal contexts play a role in visualization design, and enables us to highlight challenges and opportunities in visualization and crisis informatics research.

\subsection{Crisis-driven Changes}

Visualization changes can reflect an urgent need to adapt existing visualizations to better suit the constantly changing crisis situations; we refer to these changes as \textbf{``crisis-driven'' changes}. Existing visualizations may be insufficient to meet the challenges of visualizing crisis information as the crisis develops for a number of reasons, such as a change of data format released by health officials, or the inability of existing visualizations to reveal trends and patterns of the pandemic due to increasingly complex situations.

For example, \autoref{fig:nyt_change_over_time} shows four US maps that represent different stages of COVID-19. At the beginning of the pandemic, an early-stage map (see~\autoref{fig:nyt_change_over_time} (A)) applied dual encodings including proportional symbols (i.e., bubble map) and a qualitative thematic map (areas marked with yellow background) to show COVID-19 cases. Later on, in April, the default map was changed to a bubble map as COVID-19 cases had spread rapidly all over the country (see~\autoref{fig:nyt_change_over_time} (B)). In May, although the overall US case numbers were decreasing, some states witnessed an increasing number of cases again. To better reveal the trends and patterns, the default map changed to a choropleth map showing hot spots of how the number of new cases had changed in the last two weeks. The map classification (shown as the map legend at the top) fell into four categories: falling~\colorRect{colorLegendBlue}, about the same~\colorRect{colorLegendYellow}, rising (with three sequential colors~\colorRect{colorLegendLightOrange}\colorRect{colorLegendOrange}\colorRect{colorLegendRed}), and few or no cases~\colorRect{colorLegendLightGrey}, as shown in~\autoref{fig:nyt_change_over_time} (C). The granularity also changed from states in (A) to counties in (C) to contrast different trends in some rural areas and cites. 
A more recent version of the map classification in~\autoref{fig:nyt_change_over_time} (D) has changed to an eight-level sequential color scheme~\colorRect{colorLegendYellow}\colorRect{colorLegendNYTLevel2}\colorRect{colorLegendNYTLevel3}\colorRect{colorLegendNYTLevel4}\colorRect{colorLegendNYTLevel5}\colorRect{colorLegendNYTLevel6}\colorRect{colorLegendNYTLevel7}\colorRect{colorLegendNYTLevel8}, along with the category of few or no cases~\colorRect{colorLegendLightGrey}, since the rising second wave of cases has became the theme in August 2020.

In addition, adding the moving average lines over a time-series chart was one of the most common changes in our corpus, as shown in ~\autoref{fig:temporal} (A). Prior work has suggested that the moving average provides a more stable view of the trend than daily change and helps to convey the effectiveness of COVID-19 surveillance and containment amidst the pandemic~\cite{ng2020evaluation}. Most visualizations in our corpus that were created early on in the pandemic did not use moving averages. To our knowledge, the earliest use of moving averages was from a Singaporean report on February 29, 2020~\cite{ng2020evaluation}. Later on, more visualizations started adding moving averages to better reflect the current state of the pandemic (n=30). For example, the New York Times added the 7-day average starting from April 8, 2020. This may be because the COVID-19 case and death data started to show apparent periodic fluctuation from late March to early April.

\begin{figure*}[tb]
\centering
\includegraphics[width=\textwidth]{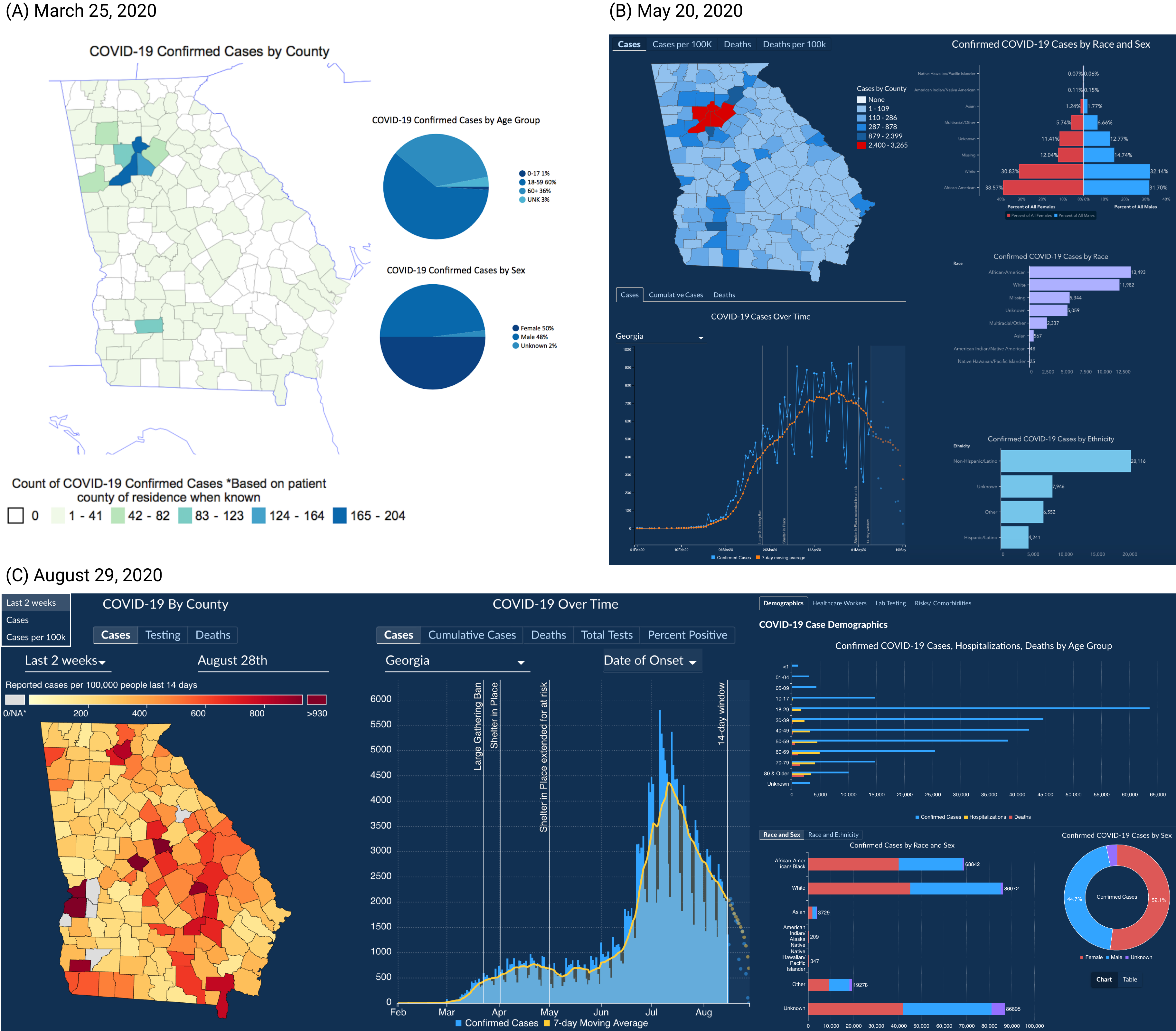}
\caption{Examples of how crisis visualizations have changed over time, by the Georgia Department of Public Health~\cite{GA_covid}:
(A) An early version of the COVID-19 dashboard (March 25, 2020) contains one (non-normalized) choropleth map and two pie charts showing the break-down COVID-19 data by age and sex;
(B) A version published in May includes choropleth maps (both showing the raw case numbers and normalized data) showing cases and deaths change over time (bottom left), and a set of bar charts showing the break-down COVID-19 information by sex, race, and ethnicity;
(C) A more recent version (August 29, 2020) provides more views to show daily status. 
}
\label{fig:GA_change_over_time}
\end{figure*}

\subsection{Public-driven Changes}

Some changes in visualizations are prompted by the public participation and critique of visualization works; these changes related to the feedback from the public and can be termed \textbf{``public-driven'' changes}. The growing interests and practices of designing and disseminating COVID-19 visualizations have led to increasing public critiques and discussions on social media (e.g., Twitter and Reddit). This sort of collective intelligence from the public has called attention to necessary changes. For instance, ~\autoref{fig:GA_misleading} (A) shows a controversial example of potentially ``manipulative'' data using ``misleading'' visualizations to show a downward trend in cases~\cite{GA_misleading}. The Georgia Department of Public Health released a graph on May 17, 2020, showing COVID-19 cases had supposedly fallen among five hardest-hit Georgia counties from April 26 to May 9, with the dates on the chart being out of order, as shown in \autoref{fig:GA_misleading} (A). Many social media users critiqued this misleading visualization and pointed out the mistakes. The graph has been taken off the website since then. A spokeswoman for Gov. Brian Kemp tweeted on May 11, 2020: \textit{``The x axis was set up that way to show descending values to more easily demonstrate peak values and counties on those dates. Our mission failed. We apologize. It is fixed.''}~\cite{GA_tweet}. The team amended the error by replacing it with the chronological order, as shown in \autoref{fig:GA_misleading} (B). The corrected chart shows that cases were holding steady rather than declining.

It is difficult to assess the intentionality of the visualization design; one could argue that the graph was manipulated to intentionally deceive or was mistakenly created by the visualization tool. Yet the effect of such visualizations may have spawned misunderstanding and misinformation that will ultimately influence people's trust with authorities and their own decision-making processes when dealing with the crisis (e.g., they underestimate the severity of the crisis). As such, this example shows how public participation and critiques of misleading visualizations can be powerful ways of bringing immediate and effective changes to visualization designs.

\subsection{Content Producer-driven Changes}

Another type of change can be framed as \textbf{``content producer-driven changes''}---changes that are mainly made in terms of the visual encodings (e.g., color and shape). For example, despite the errors they made in previously published visualizations, the Georgia Department of Public Health has also made numerous improvements in COVID-19 visualizations. \autoref{fig:GA_change_over_time} provides a snapshot of how their daily status visualizations have changed from early March to late August 2020. There are several major changes in how they have visualized COVID-19 information, including changing color schemes, normalization methods (e.g., from non-normalized choropleth maps to normalized ones), and basic chart types (e.g., from line charts to bar charts to show how case numbers have changed over time).  

\textbf{Summary:} In this section, we have presented several examples of changes being made to visualizations amidst the pandemic. Examining how visualizations have changed amidst the dynamic nature of the pandemic highlights a broader challenge in designing and evaluating crisis visualizations. That is, how can we incorporate and embrace visualization changes over time, while also being critical of and responsible in implementing such changes? We will further discuss the challenges and opportunities in the design and evaluation of crisis visualizations shortly in ~\autoref{sec:discussion}.  


\section{General Discussion}
\label{sec:discussion}

\jz{A crisis is characterized by its intense and broad impact, high risk, urgency, fast-evolving nature, and high level of uncertainty~\cite{shaluf2003review}. As we have seen with the COVID-19 pandemic, these intersecting characteristics of a crisis can lead to a large number of publicly-available visualizations, all generated from different data sources and with the great potential to collectively impact public behavior. Furthermore, COVID-19 has demonstrated how the fast-evolving nature of a crises can result in constant changes in visualization techniques and messages. In this paper, we have described four components of our conceptual framework for crisis visualizations: \textit{who}, (uses) \textit{what data}, (to communicate) \textit{what message}, \textit{in what form}, and have presented how visualizations have changed over the course of the COVID-19 pandemic. Visualization creators have updated, adapted, and transformed the crisis visualizations regularly to suit various needs, different data, and to answer ever-changing public questions. From the consumer's side, there is also a tendency to view these crisis visualizations over and over, to monitor the situation rather than to understand it once.} 

In this section, we further discuss our framework regarding its scope and utility, as well as a set of research directions drawing on our observations and analysis of existing COVID-19 crisis visualizations. 
 
\subsection{Who}  
Though we have briefly described the information outlets that have produced visualizations amidst COVID-19, we believe there is a need to further examine issues around their positioning and power (i.e., ``the current configuration of structural privilege and structural oppression in which some groups experience unearned advantages''~\cite{dignazio2020seven}). The increasing political and economic interests of news outlets and those individuals and entities who have the power to control the outlets pose challenges to neutrality in their reports~\cite{elejalde2018nature}. 
Media bias (i.e., an internal bias exhibited in media coverage)~\cite{hamborg2019automated} impacts how people perceive the risks and the severity of the crisis~\cite{slovic1987perception}. As visualizations become more accessible and popular, there is a vital need to investigate to what extent visualization designs systematically vary between different media groups (e.g., those with different political biases).  

Our findings suggest that different media outlets have shown different patterns in communicating various messages and message components (e.g., visualizations focused on socioeconomic status were all from news media). Future work should further examine these and other patterns in crisis visualization variation across information outlets. For example, how do left and right-leaning media outlets choose their color schemes for maps that seek to inform people of the severity of the crisis? How might media biases be related to visualization color choices? These questions are important to examine, as a color has different cultural meanings in different countries, and color can elicit an emotional response~\cite{welhausen2015visualizing}. Using different color schemes may influence how people perceive risks and the severity of the crisis. The importance of exploring this aspect is further supported by a recent study focused on COVID-19 visualizations~\cite{cay2020understanding}. The study shows that the choice of colors and design can greatly impact users' risk perception. Understanding the differences and similarities of visualization design across media outlets may help us examine media biases, positioning, and power issues inherent in widely distributed visualizations.

\subsection{What Data} 
Issues of crisis data sources are prevalent, as we described in ~\autoref{sec:data}. Similar to the positioning and power issues reflected in analyses of ``who'' designs the visualizations, the same concerns also extend to ``what data'' has been used for creating these visualizations. Yet, the issues surrounding data sources do not only pertain to COVID-19. Research on historical crises (e.g., SARS, H1N1, and Ebola) have already documented similar issues with data sources, such as varied data formats for reporting and dissemination~\cite{cori2017key}. Each of these threats to data quality in turn threatens the integrity of resulting visualizations. As such, care must be taken when visualizing crisis data with varied quality. For example, we emphasize the importance of reporting on the source (and even the source of source) and recency of data. Crises such as the COVID-19 pandemic introduce particular threats to the production of trustworthy visual information. Having a trustworthy data source is crucial for creating visualizations, especially in times of crisis where misinformation and disinformation are rampant. As we described in~\autoref{sec:data}, 7\% of visualizations did not cite the original data sources, leaving the audiences uncertain about where the data came from. In addition to reporting data sources, we highlight the importance of reporting the recency of data, that is, the date of data collection and visualization production. And, our discussion of how COVID-19 visualizations have changed over time, highlight how visualizations produced at one point in time may not be sufficient for depicting the state of the crisis at a later time. 

\subsection{What Message}
We have presented six high-level categories of messages that communicate information about COVID-19, including informing of severity, forecasting trends and influences, explaining the nature of the crisis, guiding risk mitigation, gauging the multifaceted impacts of the crisis, as well as communicating risk, vulnerability, and equity (\autoref{sec:messages}). Through our inductive and deductive analysis approach, we derived a categorization that paints a holistic picture of the messages within our corpus of visualizations. At the same time, given that these categories were derived from our specific COVID-19 visualization sample, future work should explore to what extent this categorization needs iteration or expansion to fit future crisis contexts. 

\jz{By categorizing the messages conveyed in COVID-19 visualizations, we aim to showcase their diversity and catalyze research that evaluates them and their impact. Our analysis shows a large spectrum of crisis-related messages being conveyed and many visualizations used for each message. While crisis and non-crisis public health visualizations may convey some similar messages, COVID-19 provides a unique opportunity to develop the message categorization we present in this paper. Such categorizations are more challenging in other public health contexts where there are fewer visualizations produced. 
In addition, some messages are especially salient in crisis visualizations. For example, these communication tools have a particular focus on helping people gauge the crises' impact, given the intense and broad influences that they can have on the world. As another example, when communicating severity, risk, and equity, crisis visualizations may especially highlight medical resources, since such resources are especially crucial in these contexts. Also, the amount of visualizations communicating each message fluctuates as crises evolve. For instance, as public interests shift, we see more COVID-19 visualizations have turned to conveying felt experience and acknowledging grief and uncertainty.} 

Moreover, our categorization of messages (all starting with verbs) provides a high-level task analysis that can be helpful to guide task abstraction and visualization design. Conducting task abstraction is based on task analysis. The goal of task abstraction is to recast user tasks and goals from domain-specific languages to a generalized terminology to allow for better understanding and readability of domain tasks~\cite{brehmer2013multi}.  For example, building upon our categorization of high-level messages and tasks for crisis visualizations, future work could consider using existing task abstraction frameworks, ranging from high-level~\cite{shneiderman1996eyes} to low-level abstraction approaches~\cite{amar2005low}, or works that aim to bridge between the high- and low-level abstraction (e.g.,~\cite{brehmer2013multi, zhang2019idmvis}). Our work represents an important first step towards categorizing a set of tasks, or goals, for crisis visualizations, which will help future work create task abstraction that can further guide the choice of crisis visualization forms. \jz{General guidelines may be largely applicable to visualization design amidst a crisis. However,  additional considerations may be needed to suit specific crisis contexts; our framework and analysis can help support the building of such guidelines, by focusing attention on the unique features of crises.}

\subsection{In What Form}
We have characterized patterns and trends in our corpus as well as challenges inherent in the design of such crisis visualizations (see \autoref{sec:visual_forms}).  
Our analysis provides both a horizontal overview that compares and contrasts across visualizations aiming to communicate the same type of messages (even with the same data source), as well as a vertical overview that examines how visualizations have changed over time. Our findings can help to bootstrap the creation of the next generation of crisis visualizations, by characterizing previously used visual techniques along with the potential pitfalls of such approaches. These findings can help future visualization designers to avoid reinventing the wheel, thus accelerating innovation and the more efficient production of visualizations that communicate crisis information. 
 
As the COVID-19 pandemic evolves, there may be new opportunities and imperatives to develop novel visualization methods that help people make sense of trends and patterns within the pandemic. Yet, using new visualization methods may make it challenging for people to interpret the message accurately. For example, using a log scale may help better display a trend. However, one study found that people had a less accurate understanding of the trajectory of the pandemic when showed the number of deaths on a log scale~\cite{romano2020covid}. Therefore, care must be taken when using novel visualization techniques or approaches that the general public is less familiar with.  

However, we should also think critically regarding the production of crisis visualizations. The great diversity and the number of visualizations in our corpus reflect an increasing public interest in visualization. Yet, this proliferation of visualizations has also led to issues such as the varied quality among the visualizations reflected in our corpus. Thinking back to the ``misleading'' visualization example from the Georgia Department of Public Health (\autoref{fig:GA_misleading}), the website has been visited 4.93 million times according to Similarweb (a website traffic tracker)~\cite{similarweb}. We cannot say for certain that the chart was created intentionally or unintentionally (e.g., due to lack of knowledge or technical mistake). But these charts may have caused negative consequences to people who had seen them (e.g., not following public health recommendations). However, such ``unintended consequences'' may be able to be anticipated in advance~\cite{parvin2020unintended}. Neglecting such unintended consequences may cause more serious issues and may become a way to marginalize the ethical questions at the root of design decisions. 

Therefore, instead of advocating for a single yes or no response to the question of whether or not we should encourage more crisis visualizations, why not think about, for example, how we can design more effective crisis visualizations through collaborations between experts in design, public health, and crisis informatics? Though our work discusses a set of visualization techniques used, issues, and challenges in the context of crisis visualizations, more work is needed to build a more comprehensive glossary in this field. For example, building upon our work, future work can compile a list of visual components, encodings, and visualisation techniques that are specifically related to the crisis context. Such glossaries and visualization ``cheatsheets'' would be helpful for people who typically are under time pressure during crises, to create crisis visualizations more effectively and efficiently.

\subsection{Under What Circumstances}
Our findings show how visualizations have changed over time, including crisis-driven, public-driven, and content producer-driven changes. Our exploration of the dynamic temporal contexts in crisis visualizations offers a novel way to reflect on history, as visualization changes over time offer a means of reflecting on crisis events. Such a record of crisis visualizations and reflection upon them can be seen as a way of building \textit{cultural heritage} (i.e., artifacts and memories that help to characterize our society's experience of the COVID-19 pandemic), which in this context, can be seen as building up and examining our sense of cultural identity. Exploring cultural heritage involves ``renegotiating our identities and value systems by reworking the meanings of the past as the cultural, social and political needs of the present change''~\cite{liu2010grassroots}. As such, studies of visualization changes over time can be an important cultural reflection process, and we hope that our collection of visualizations, analysis of them, and the resulting framework can help to catalyze the preservation of important cultural records and future reflections on these records. On the one hand, visualizations need to be adapted to better present and communicate the constantly changing information as a crisis develops. On the other hand, after decades have passed, the documented changes of visualizations can help us better recall how the crisis developed ``back then''.   
 
Moreover, as we mentioned before, the general public on social media participated in detecting, criticizing, and redesigning existing visualizations, including those ``misleading'' visualizations. Their collective critiques and redesign of these crisis visualizations---emerging from collective knowledge building---contributes a collective memory that may shape the form of cultural heritage. Scholars have pointed out that research examining how memory-focused technologies can operate at a collective and cultural level is an underexplored field ~\cite{liu2010grassroots, van2008technologies}. The ``grassroots heritage'' framework proposed by Liu~\cite{liu2010grassroots} offers inspiration for exploring the temporal aspects of visualizations (e.g., how these visualizations change, and why). For example, some concepts described in the grassroots heritage framework include an open collection of content on a societal scale, and fostering interpretation of these collections, which are in line with the scope of our work.  

People are increasingly engaged with visualization design in times of crisis using social media. As such, \textit{people are changing culture as they participate}. We encourage future work that explores the public's engagement with crisis visualizations, the resulting changes to those visualizations, and how such participation in the building of cultural heritage amidst a pandemic might impact the public (e.g., by helping or hindering individuals' ability to cope with and make sense of the crisis).

\subsection{To Whom} 
\textit{``To whom''} refers to the receiver of the message or an audience~\cite{lasswell1948structure}. People use visualization to communicate a message. A challenge of communication during a crisis is that not all populations can be reached effectively and equally~\cite{reynolds2007cdc}. It is more challenging for special populations to assess information due to various constraints, such as cognitive, physical impairment or \jz{sensory impairment~\cite{COVID_visual_impaired, Holloway20Nonvisual}}, language barriers, and lack of devices~\cite{reynolds2007cdc}. The effects of crisis communication through visualizations also depend on the demographics, literacy, numeracy, and personal traits of the audiences~\cite{borner2019data}. Moreover, the relationship to \textit{``what message''} and \textit{``who''} also plays a role in effecting visualization communication, due to pre-existing confirmation biases (e.g., strong distrust of organizations distributing messages or visualizations)~\cite{reynolds2005crisis, van2019communicating, zhang2020ci}. All these factors associated with the characteristics of the audiences and the relationship and trust between the targeted audiences and information outlets should be considered when tailoring visualization design and dissemination in crisis communication. Future work can make use of our categorization of messages and example visualizations to investigate how crisis visualizations are perceived by different populations.
\vspace{-5pt}

\subsection{With What Effect}
Though our work did not examine the effects of crisis visualizations generated in times of COVID-19 on the general public, there is a critical need to evaluate such effects. Indeed, a great body of prior work has explored the effects of visualization on various aspects of people's risk perception and decision-making process~\cite{lipkus2007numeric}. While our findings show that there have been many variations of emerging novel visualization encodings and techniques raised amidst the pandemic, there is still a need to further examine how visualization impacts people. 
Our work provides a collection of COVID-19 visualizations and a basic framework to accelerate future work to examine how various visualization encodings and techniques within each type of message impacts people's cognition, emotion, trust, behaviors, mental health, and decision-making processes. Understanding these effects will help us design and communicate crisis information more effectively in the future.  

In addition to the traditional usability and performance metrics for visualization evaluation (e.g., accuracy, completion time, and memorability), we suggest future work explore dimensions of the felt and lived experience. The examples of data-driven storytelling approaches of acknowledging grief and misery of the loss (described in \autoref{sec:inform_severity} may spark inspiration for future evaluation studies. Our notion is in line with some visualization scholars~\cite{boy2017showing, kennedy2018feeling, peck2019data, saket2016beyond} that also indicated the importance of paying attention to the impact of emotion in visualization. Yet, we also argue that there is more to explore regarding various aspects of assessing felt and lived experiences ``in the wild'' especially in times of crisis, where uncertainty, negative emotions, and distrust are rampant. 

Moreover, the effect of dynamic temporal contexts in crisis visualization design brings up a new challenge in evaluation. Since visualizations have changed over time, what methodological approaches should we take into consideration when designing and evaluating visualizations? In addition to traditional considerations like whom do we evaluate? How many participants do we need? What do we evaluate? Other considerations include: what are the metrics we should consider to incorporate the waves of changes in visualizations over time? Shneiderman and Plaisant's Multi-dimensional In-depth Long-term Case studies (MILCs) might help to examine the long-term effects of visualizations~\cite{shneiderman2006strategies}. However, there is more work needed to further explore methodological and technical aspects of visualization evaluation ``in the wild'' during times of crisis.   

It is also important to understand the similarities and differences of various aspects of effects among populations with different characteristics and beliefs in relationship and trust with authorities and organizations (as discussed in the \textit{``to whom''} component). Examining these aspects is crucial for creating an equitable society in which people of all backgrounds are able to utilize, benefit from, question, and challenge information that is disseminated about the humanitarian challenges that impact our world.

\section{Limitations}
Our work was limited by the sample of COVID-19 visualizations, not covering other historical crises. Therefore, more research is needed to test our framework and adapt it to better suit contexts of the crises. Moreover, our collection inevitably fails to capture all COVID-19 visualizations. Most of the examples (in English) in this paper were from publishers in the United States, and some were from other countries (e.g., Japan, Singapore). These examples in the paper may create bias as they may be seen as Western-centric, but they did not reflect our intention. Instead, we hope our work will spark further conversations around crisis visualizations. 

\section{Conclusion}
Not a single visualization or message can tell the whole story about the pandemic. Instead, multiple views across different fields with a collective effort from multiple stakeholders may help better reveal the state of the crisis. Through our work in collecting, documenting, organizing, and analyzing hundreds of COVID-19 visualizations, we categorized the trends and patterns of these crisis visualizations, and challenges inherent in the design. We also contributed a conceptual framework of crisis visualization that helps guide future analysis of existing crisis visualization, design effective visualizations, and evaluate the effects of these visualizations. Our work aims to help future work further examine the space of crisis visualization. 

\begin{acks}   
We wish to thank Ben Shneiderman, Catherine Plaisant, John Stasko, and Alex Endert for discussions and advice, and our reviewers for their constructive feedback. We also thank Paul Kahn for co-leading the database and people who have facilitated with data collection, as well as Lin Shi, Jennifer Howell, Rumi Chunara, Racquel Fygenson, Helia Hossein-pour, Anamaria Crisan, Hugh Dubberly and Dubberly Design Office, Wellness Technology lab at Georgia Tech, Visualization group at Georgia Tech, and VIS lab at Northeastern University for feedback and various levels of support on this work. This work was supported by NSF award number \#2028374. 
\end{acks}

\balance

\bibliographystyle{ACM-Reference-Format}
\bibliography{ref}

\end{document}